\documentclass[aps,nofootinbib,11pt,preprintnumbers]{revtex4}
\usepackage{amssymb,amsmath}
\usepackage{hyperref}
\usepackage{graphics}
\usepackage{graphicx}
\newcommand{\bea}{\begin{eqnarray}}
\newcommand{\eea}{\end{eqnarray}}
\newcommand{\beq}{\begin{equation}}
\newcommand{\eeq}{\end{equation}}

\newcommand{\eq}[1]{Eq.~(\ref{#1})}
\newcommand{\eqs}[1]{Eqs.~(\ref{#1})}
\newcommand{\ave}[1]{\langle {#1} \rangle}

\newcommand{\lb}{\bar{\ell}}
\def\x{{\bf x}}
\def\tmu{\tilde{\mu}}
\def\E{{\bf E}}
\def\F{{\bf F}}
\def\K{{\bf K}}
\def\0{{\bf 0}}

\def\p{{\bf p}}

\def\nave{{\bar n}}

\begin{document}
\preprint{INT-PUB-10-033}

\title{
Influence of vector interaction and Polyakov loop dynamics on inhomogeneous 
chiral symmetry breaking phases
}

\author{Stefano~Carignano$^1$}
\author{Dominik Nickel$^2$}
\author{Michael~Buballa$^1$}

\affiliation{$^1$Institut f\"ur Kernphysik, Technische Universit\"at Darmstadt, Germany\\
$^2$Institute for Nuclear Theory, University of Washington, Seattle, USA
}

\date{July 2010}

\begin{abstract}
\noindent
We investigate the role of the isoscalar vector interaction and the 
dynamics of the Polyakov loop on inhomogeneous phases in the phase diagram of 
the two-flavor Nambu-Jona--Lasinio (NJL) model.
Thereby we concentrate on inhomogeneous phases with a one-dimensional 
modulation, explicitly domain-wall solitons and, for comparison,
the chiral spiral.
While the inclusion of the Polyakov loop merely leads to quantitative changes 
compared to the original NJL model, the inclusion of a repulsive 
vector-channel interaction has significant qualitative effects:
Whereas for homogeneous phases the first-order phase transition gets weakened 
and eventually turns into a second-order transition or a cross-over, the domain of
inhomogeneous phases is less affected. 
In particular the location of the Lifshitz point in terms of temperature and density is 
not modified.
Consequently, the critical point disappears from the phase diagram and 
only a Lifshitz point (showing a different critical behavior) remains.
In particular, susceptibilities remain finite.
\end{abstract}

\maketitle

\section{Introduction}

\noindent
Due to its potentially rich and complex structure a better understanding of 
the phase diagram of quantum chromodynamics (QCD) still poses one of the 
biggest challenges in modern nuclear physics (for dedicated reviews see, e.g., 
Refs.~\cite{Halasz:1998qr,Stephanov,BraunMunzinger:2008tz}
). Experimentally, the domain of large temperatures is explored by heavy-ion 
collisions whereas the properties at low temperatures and large densities are 
relevant for compact stellar objects. 
On the theoretical side ab-initio lattice calculations are limited to small 
chemical potentials and investigations at moderate densities so far mainly 
rely on phenomenological models.
Probably the most widely-used one in this context is the Nambu-Jona--Lasinio 
(NJL) model~\cite{NJL} and its extensions, sharing global symmetries as well 
as the phenomenon of chiral symmetry breaking with QCD
(see Refs.~\cite{Vogl:1991qt,Klevansky:1992qe,Hatsuda:1994pi,Buballa:2005} 
for reviews).

\noindent
In this article we explore the role of inhomogeneous phases in the phase 
diagram of NJL-type models, focusing on the inclusion of a vector-channel 
interaction as well as the Polyakov-loop dynamics.
The isoscalar-vector interaction naturally arises in the NJL model 
when motivating the interaction from a one-gluon exchange in 
QCD~\cite{Vogl:1991qt}
and was already shown to be of particular importance in the Walecka model at 
finite densities~\cite{Walecka:1974qa}.
More recently, its influence on the location and emergence of critical 
points in the phase diagram has attracted new 
interest~\cite{Fukushima:2008,Fukushima:2008b,Zhang:2009mk}.

\noindent
The coupling of the NJL model to the Polyakov-loop dynamics on the other hand 
has been introduced to mimic features of (de-)confinement, in particular at 
finite temperatures and vanishing 
densities~\cite{Fukushima:2003fw,Ratti:2005jh}.
There are several open issues which complicate the setup of the model at
finite chemical potential~\cite{Schaefer:2007pw,Abuki:2008iv}.
However, at vanishing temperatures, the Polyakov loop always decouples 
from the quark sector by construction. 
In the regime of low temperatures and (not too) high chemical potential
one typically finds a chirally restored but ``confined'' phase, which
is sometimes related to the quarkyonic phase~\cite{McLerran:2008ua}, suggested 
in Ref.~\cite{McLerran:2007qj} for QCD in the limit of a large number of colors
($N_c$).

\noindent
Although most studies are usually restricted to homogeneous phases, 
the importance of chiral crystalline phases being characterized by an 
inhomogeneous order parameter has been pointed out long ago.
Well known examples are the Skyrme crystal~\cite{Goldhaber:1987pb}
and the chiral density wave (CDW)~\cite{Broniowski:1990dy}.
For quark matter, chiral crystalline phases have been explored in the weakly 
coupled regime for large $N_c$~\cite{Deryagin:1992rw,
Shuster:1999tn}, where they form the ground state at vanishing temperatures 
even at asymptotic densities. Related to that they naturally show up in the 
quarkyonic matter picture~\cite{Kojo:2009ha}
and in holographic models~\cite{Rozali:2007rx}.
For metals, however, the underlying mechanism of  particle-hole pairing was however already considered in the 1960s~\cite{Overhauser:1962}. 
In a similar context inhomogeneous phases have also been discussed for
color superconductivity, see e.g. Refs.~\cite{Alford:2000ze,
Bowers:2002xr,Casalbuoni:2005zp,Mannarelli:2006fy,Rajagopal:2006ig,
Nickel:2008ng,Sedrakian:2009kb}, again borrowing ideas from condensed matter physics~\cite{Fulde:1964zz,LO64}.

\noindent
An inhomogeneous phase also exists in the large $N_c$-limit of the
$1+1$-dimensional Gross-Neveu (GN) model~\cite{Gross:1974jv},
where one has a rather complete picture of the phase 
diagram~\cite{Schnetz:2004vr,Schnetz:2005ih,Thies:2006ti,Basar:2009fg}.
Here the chiral crystalline phase emerges at low densities by formation of kink and antikink solitons and reaches out to arbitrarily high densities for sufficiently small temperatures.

\noindent
In the NJL model, most investigations of inhomogeneous phases have been
performed for the CDW (``chiral spiral''), corresponding to a single
plane wave ansatz for the order parameter~\cite{Sadzikowski:2000ap,
Nakano:2004cd}. The notable exception here is Ref.~\cite{Rapp:2000zd}. More recently, it was shown that the known solutions
from the $1+1$-dimensional GN model can be used to construct solutions
of the $3+1$ dimensional NJL model whose order parameter varies in one spatial 
direction~\cite{Nickel:2009wj}.
These solutions are more favored than the chiral spirals and 
the corresponding phase occupies a larger region of the phase diagram.
However, due to the short-ranged interaction (in contrast to the 
weak-coupling QCD analysis at large $N_c$) and different kinematics 
compared to the GN model, this phase is constrained to a finite range 
of densities.
 
\noindent
In the chiral limit, we can thus distinguish three different phases,
namely the homogeneous chirally broken phase, the inhomogeneous phase,
and the chirally restored phase.
These phases meet in a single point, a so-called ``Lifshitz point'' 
(LP). 
Within a Ginzburg-Landau analysis it can be shown that in the NJL model 
without vector interactions the LP exactly coincides with the (tri-) critical 
point (CP) of the chiral phase transition, which would be present if the 
analysis was limited homogeneous phases~\cite{Nickel:2009ke}. 
In Ref.~\cite{Nickel:2009wj} this was confirmed by an explicit model
calculation. 
Moreover, it was found that the inhomogeneous phase is bordered by 
second-order phase boundaries, and that it completely covers the
would-be first-order transition line between the homogeneous phases.

\noindent
In the present work we extend these investigations to include repulsive
vector interactions.
As the main result we find that the LP stays at the same temperature,
whereas, when the analysis is restricted to homogeneous phases, 
the CP is shifted to smaller temperatures and larger chemical potentials.
The critical region surrounding the CP thus disappears from the phase diagram as it is replaced 
by an energetically more preferred inhomogeneous ground state.
Consequently, the divergence of susceptibilities at the CP,
which has been related to event-by-event fluctuations in heavy-ion collisions
and discussed as one of the most important experimental signatures of the 
CP~\cite{Stephanov:1998dy}, is removed from the phase diagram.
For completeness, we also discuss the inclusion of the Polyakov-loop dynamics, which affects the structure of the phase diagram only quantitatively.

\noindent
This paper is organized as follows: 
In section~\ref{sec:NJLvector} we discuss the role of the isoscalar-vector interaction when investigating the NJL model's phase diagram allowing for inhomogeneous phases. After introducing the model and the general framework for investigating inhomogeneous phases in subsections~\ref{subsec:model} and~\ref{subsec:onedimensionalmods}, respectively, we move on to an extensive numerical study of the phase diagram in subsection~\ref{subsec:numGv}. The latter includes results for the density profiles in solitonic ground states, a comparison to the CDW, a discussion of number susceptibilities and the role of finite current quark masses. Part of the obtained results is then explained in the context of a generalized Ginzburg-Landau (GL) expansion in subsection~\ref{sec:GL}.
The subsequent section~\ref{sec:Polyakov} is devoted to the inclusion of Polyakov loop dynamics and its effect on the structure of the phase diagram.
Finally, we summarize and conclude in section \ref{sec:discussion}.

\section{The two-flavor NJL model with vector-channel interaction}
\label{sec:NJLvector}

\subsection{Model and Approximations}
\label{subsec:model}
\noindent
In view of finite density investigations, an important extension of the 
original NJL model is given by the inclusion of an isoscalar vector-channel 
interaction. The Lagrangian is then given by
\bea
\mathcal{L} =
 \bar{\psi}\left(i\gamma^\mu \partial_\mu - m\right)\psi +
G_S\left(\left(\bar{\psi}\psi\right)^2+\left(\bar{\psi}i\gamma^5\tau^a\psi\right)^2\right)
-
G_V  \left(\bar{\psi}\gamma^\mu\psi\right)^2
\,,
\eea
where  $\psi$ is a $4N_f N_c$-dimensional quark spinor for $N_f=2$ flavors and $N_c=3$ colors, $\gamma^\mu$ and $\tau^a$ are Dirac and Pauli matrices,
respectively, and $m$ is the degenerate current quark mass.

\noindent
In mean-field approximation we expand around the expectation values of the 
non-vanishing, potentially spatially varying scalar condensate
$S(\x)\equiv\langle\bar{\psi}(\x)\psi(\x)\rangle$, pseudo-scalar condensate 
$P_3(\x)\equiv\langle\bar{\psi}(\x)\,i\gamma^5\tau^3\psi(\x)\rangle$ and density
$n(\x)\equiv\langle\psi^\dagger(\x)\psi(\x)\rangle$, yielding
\bea
(\bar{\psi}\psi)^2 
&\simeq&
-S(\x)^2 + 2S(\x) \bar{\psi}\psi
\,,
\nonumber\\
(\bar{\psi}i\gamma^5\tau^a\psi)^2 
&\simeq&
-P(\x)^2 + 2P(\x) \bar{\psi}i\gamma^5\tau^3\psi
\,,
\nonumber\\
(\bar{\psi}\gamma^\mu\psi)^2
&\simeq&
-n(\x)^2 + 2n(\x) \bar{\psi}\gamma^0\psi
\,.
\eea
In the case of a periodic condensate with Wigner-Seitz cell $V$ and using 
the imaginary-time formalism, 
the mean-field thermodynamic potential is then given by
\bea
\Omega
&=&
-\frac{T}{V}\ln 
\int \mathcal{D}\bar{\psi}\mathcal{D}\psi \exp\left(\int_{x\in [0,\frac{1}{T}]\times V} (\mathcal{L}_{\rm mean-field}+\mu \bar{\psi}\gamma^0 \psi)\right)
\nonumber\\
&=&
\Omega_{\rm kinetic}
+
\Omega_{\rm cond}
\,,
\eea
with the individual contributions
\bea
\label{eq:Omega1}
\Omega_{\rm kinetic}
&=&
-
\frac{T}{V}
\sum_n
{\rm Tr}_{D,c,f,V}\,\ln\left(\frac{1}{T}\left(i\omega_n+H-\mu\right)\right)
\,,
\nonumber\\
\Omega_{\rm cond}
&=&
\frac{1}{V}
\int \!d\x
\left(
\frac{\vert M(\x)-m\vert^2}{4G_S}
-
\frac{(\tilde{\mu}(\x)-\mu)^2}{4G_V}
\right)
\,,
\eea
the quasi-particle Hamiltonian
\bea
\label{eq:hamiltonian0}
H-\mu
&=&
\underbrace{
-i\gamma^0\gamma^{i}\partial_{i} 
+
\frac{\gamma^0}{2}\left(M(\x)+M(\x)^* + \gamma^5\tau^3M(\x) -\gamma^5\tau^3M(\x)^*\right)
}_{\equiv H_0}
-
\tilde{\mu}(\x)
\,,
\eea
the spatially dependent ``constituent quark'' mass 
\bea
    M(\x) = m-2G_s(S(\x)+iP(\x))\,,
\eea 
the renormalized quark chemical potential 
\bea
\label{eq:tildemu}
\tilde{\mu}(\x)=\mu-2G_V\,n(\x)\,,
\eea
and Matsubara frequencies $\omega_n=(2n+1)\pi T$.

\noindent
For general spatially dependent mean fields, the evaluation of the
functional trace in \eq{eq:Omega1} and the subsequent minimization 
of $\Omega$ is highly non-trivial.
However, as shown in Ref.~\cite{Nickel:2009wj}, for the case without 
vector interactions a solution can be found if only one-dimensional 
mass modulations are considered. 
In the present work we want to generalize these solutions to $G_V > 0$.
Since this cannot be done exactly in a straightforward manner,
we approximate the density in \eq{eq:tildemu} by its spatial average,
\beq
\label{eq:napprox}
 n(\x) \;\rightarrow\; \nave \;\equiv\; \ave{n(\x)} \;=\; {\rm const}.
\eeq
As a consequence, $\tilde\mu$ becomes constant as well, and the problem
reduces to the known case without vector interaction at a shifted value
of the chemical potential. 
Of course, at first sight, it seems rather questionable whether the 
replacement \eq{eq:n} is a good approximation in an inhomogeneous phase.
However, as we will discuss in more detail later on, it can be 
rigorously justified in the vicinity of a second-order phase boundary
to the restored phase, and in particular for the Lifshitz point. 

\noindent
The kinetic contribution $\Omega_{\rm kinetic}$ to the thermodynamic 
potential can now be evaluated from the eigenvalue spectrum $\{E_i\}$ 
of the Hamiltonian $H_0$ through the relation
\bea
T\sum_n \ln\left(\frac{1}{T}(i\omega_n+E_i)\right)
&=&
\frac{1}{2}E_i+ T \ln\left(1+\exp\left(-\frac{E_i}{T}\right)\right)
\,.
\eea
Introducing a density of states to express 
$-\frac{1}{2V}\sum_{E_i}\rightarrow -2N_c \int dE\, {\rho}(E)$, 
we then formally obtain
\bea
\label{eq:Omegaf1}
\Omega_{\rm kinetic}
&=&
-2N_c 
\int\limits_0^\infty dE\, {\rho}(E)\,  f(E)
\,,
\eea
with
\bea
\label{eq:Omegaf2}
f(E)
&=&
f_{\rm UV}(E)
+
f_{\rm thermal}(E)
\,,
\nonumber\\
f_{\rm UV}(E)
&=&
E
\,,
\nonumber\\
f_{\rm thermal}(E)
&=&
T\,\ln\left(1+\exp\left(-\frac{E-\tilde{\mu}}{T}\right)\right)
+
T\,\ln\left(1+\exp\left(-\frac{E+\tilde{\mu}}{T}\right)\right)
\,,
\eea
where we have used the fact that the eigenvalues of $H_0$ come in pairs
$\{E_i,-E_i\}$.

\noindent
The last missing step is a regularization of the diverging integration. For 
homogeneous phases this is mostly done by a momentum cutoff (see, e.g., 
Refs.~\cite{Vogl:1991qt,Klevansky:1992qe,Hatsuda:1994pi,Buballa:2005}).
This is however not possible for inhomogeneous phases, since the 
quasi-particle energies can no longer be labelled by a conserved 
three-momentum. 
Instead we have to apply a regularization of the functional logarithm, e.g., 
by a proper-time regularization, 
which essentially acts on the energy spectrum rather than the quasi-particle momenta.
With the density of states growing like $\rho(E)={E^2}/\pi^2+O(E^0)$ in three 
dimensions, we then only have to regularize the contribution at 
$T=\tilde{\mu}=0$ stemming from $f_{\rm UV}(E)$. We regularize this part by a 
specific blocking function in the proper-time integral leading to a 
Pauli-Villars regularization of the form~\cite{Klevansky:1992qe}
\bea
f_{\text{UV}}(E)
\;\rightarrow\;
f_{\text{PV}}(E)
\;=\;
\sum_{j=0}^{3}c_j\sqrt{E^2+j\Lambda^2}
\,,
\eea
with $c_0=1$, $c_1=-3$, $c_2=3$, $c_3=-1$ and a cutoff scale $\Lambda$.

\noindent
Within this setup we aim to discuss ground states by considering the stationary conditions
\bea
\label{eq:stationary}
\frac{\delta \Omega}{\delta M(\x)}
&=&0
\,,
\\
\label{eq:stationary2}
\frac{\partial\Omega}{\partial \tilde{\mu}}&=&
0
\,,
\eea
\noindent
determining the order parameter $M(\x)$ and the shifted chemical
potential $\tilde{\mu}$. We recall that the latter is a space-independent
constant after inserting our approximation \eq{eq:napprox}
into \eq{eq:tildemu}.
For the later discussion it is helpful to cast condition~(\ref{eq:stationary2}) 
into
\bea
\label{eq:gapeqmueff}
\mu
&=&
\tilde{\mu}
+
2G_V\,\nave
\,,
\eea
where the average density is given by
\bea
\label{eq:n}
\nave
&=&
2N_c\int\limits_0^\infty\!dE\,{\rho}(E)(n_+ - n_-) 
\eea
with the occupation numbers $n_\pm=1/(1+\exp((E\mp\tilde{\mu})/T))$.
We emphasize that, besides explicitly depending on $\tilde\mu$ via the 
occupation numbers,
$\nave$ is a functional of $M(\x)$ via the density of states $\rho(E)$.

\noindent
At this stage it is worth noting that for most quantities
all dependence on $G_V$ can be absorbed into $\tilde{\mu}$. 
In particular the form of the gap equation (\ref{eq:stationary}) and,
hence, its solutions $M(\x)$ are identical to the case without  
vector interaction upon replacing $\mu\rightarrow\tilde{\mu}$.
As obvious from \eq{eq:n},
the same is true for the average density $\nave$.
As a consequence, the mass functions $M(\x)$ at a given $\nave$
do not depend on $G_V$.
For homogeneous phases this is a well-known result~\cite{Buballa:2005}.
The remaining effect of $G_V$ is on the one hand side to map $\tilde{\mu}$ onto 
$\mu$ via Eq.~(\ref{eq:gapeqmueff}), 
and on the other hand to shift the value of the thermodynamic potential 
of a solution by $-G_V \nave^2$.
This will be important for the explanation of our results later on.

\subsection{One-dimensional modulations}
\label{subsec:onedimensionalmods}

\noindent
For the explicit evaluation of the expressions derived in 
Sec.~\ref{subsec:model} we still need to determine the density of states
$\rho(E)$. In general, for arbitrary mass functions $M(\x)$,  
this is highly non-trivial and the problem gets even more
involved for the subsequent variation in $M(\x)$.
However, as mentioned earlier, if we restrict ourselves to 
phases with one-dimensional modulations the task can be reduced to a problem 
in the $1+1$-dimensional GN model~\cite{Nickel:2009wj}.
Since furthermore for the GN model all inhomogeneous phases have been 
classified~\cite{Correa:2009xa} and in particular the phase diagram has been 
discussed in 
detail~\cite{Schnetz:2004vr,Schnetz:2005ih,Thies:2006ti,Basar:2009fg}, 
the investigation of these phases simplifies strongly.

\noindent
In the following we always assume that the one-dimensional modulations
are in $z$-direction, i.e., the system is invariant under translations
in the $xy$-directions. 
In the chiral limit the favored mass functions then take the 
form\footnote{This expression can be rewritten as
$M_{\rm soliton}(\x) = \sqrt{\nu'}\Delta'\,{\rm sn}(\Delta' z\vert\nu')$
where $\nu' = (\frac{1-\sqrt{1-\nu}}{1+\sqrt{1-\nu}})^2$
and $\Delta' = (1+\sqrt{1-\nu})\Delta$, but \eq{eq:Mz} is
more convenient for our purpose.
}
\beq
\label{eq:Mz}
M_{\rm soliton}(\x)
=
\nu\Delta\,\frac{ {\rm sn}(\Delta z\vert\nu) {\rm cn}(\Delta z\vert\nu)}{{\rm dn}(\Delta z\vert\nu)}
\,,
\eeq
where ${\rm sn}$, ${\rm cn}$, and  ${\rm dn}$
are Jacobi elliptic functions, and $\nu$ and $\Delta$ are two independent 
parameters.
$M_{\rm soliton}(\x)$ parameterizes a lattice of domain-wall solitons. 
For $\nu\rightarrow 1$ this solution becomes thermodynamically degenerate 
with a homogeneous phase, since this limit 
corresponds to a single soliton localized around $z=0$, 
which does not contribute in the thermodynamic limit.

\noindent
The density of states in this phase is explicitly given by~\cite{Nickel:2009wj}
\bea
\label{eq:rho}
{\rho}_{\rm soliton}(E)
=
\frac{E\Delta}{\pi^2}\left\{\phantom{\frac{\E}{\K}}\right.\hspace{-5mm}
&&
\theta(\sqrt{\tilde{\nu}}\Delta-E)
\left[
 \E(\tilde{\theta} \vert\tilde{\nu})+\left(\frac{\E(\nu)}{\K(\nu)}-1\right) 
\F(\tilde{\theta} \vert\tilde{\nu})
\right]
\nonumber\\
&+&
\theta(E-\sqrt{\tilde{\nu}}\Delta)
\theta(\Delta-E)
\left[
 \E(\tilde{\nu})+\left(\frac{\E(\nu)}{\K(\nu)}-1\right) \K(\tilde{\nu})
\right]
\nonumber\\
&+&
\left.
\theta(E-\Delta)
\left[
\E(\theta \vert\tilde{\nu})
+\left(\frac{\E(\nu)}{\K(\nu)} -1\right)\F(\theta \vert\tilde{\nu})
+\frac{\sqrt{(E^2 - \Delta^2) (E^2 - \tilde{\nu}\Delta^2)}}{E\Delta}
\right]
\right\}
\,,
\nonumber\\
\eea
where {\K} and {\F} are the complete and incomplete elliptic integrals of 1st
kind, respectively, and {\E} are the (complete or incomplete) elliptic 
integrals of 2nd kind.
Furthermore we introduced the notations
$\tilde{\nu}=1-\nu$, $\tilde{\theta}=\arcsin(E/(\sqrt{\tilde{\nu}}\Delta))$, 
and $\theta=\arcsin(\Delta/E)$.

\noindent
With these expressions at hand, the stationary condition~(\ref{eq:stationary})
for the ground state of the system is reduced from a complicated
functional derivative to the much simpler problem of 
extremizing the thermodynamic potential in the two parameters
$\Delta$ and $\nu$.
As argued in Ref.~\cite{Nickel:2009wj} the solutions obtained by extremizing the thermodynamic 
potential in the remaining parameters then also fulfill the more
general stationary condition~(\ref{eq:stationary}).

\noindent
We can also consider finite current quark masses, for which the order 
parameter is generalized to
\bea
\label{eq:MzReal}
M_{\rm soliton, m}(z)
&=&
\Delta
\left(
\nu\,
\mathrm{sn}(b\vert \nu)
\mathrm{sn}(\Delta z\vert \nu)
\mathrm{sn}(\Delta z+b\vert \nu)
+
\frac{
\mathrm{cn}(b\vert \nu)\mathrm{dn}(b\vert \nu)
}{
\mathrm{sn}(b\vert \nu)
}
\right)
\,.
\eea
Here $b$ is an additional parameter to be varied together with $\Delta$ and
$\nu$. The density of states changes to
\bea
{\rho}_{\rm soliton,m}(E)
&=&
\frac{E}{\sqrt{E^2-\delta \Delta^2}}{\rho}_{\rm soliton}
(\sqrt{E^2-\delta \Delta^2})\theta(E-\sqrt{\delta}\Delta)\,,
\eea
where $\delta\in [0,\infty]$ is given by 
$\mathrm{sn}(b\vert \nu) = \frac{1}{\sqrt{1+\delta}}$. The chiral limit, i.e. $m=0$, corresponds to $\delta=0$ or equivalently $b=\K(\nu)$.

\noindent
For comparison we will sometimes also consider the chiral density 
wave (CDW or ``chiral spiral'') defined by the mass function 
\beq
\label{eq:MCDW}
M_{\rm CDW}(\x)
=
\Delta \exp(i q z)
\,.
\eeq
In this case we restrict ourselves to the chiral limit and
the corresponding density of states is given by~\cite{Nickel:2009wj}
\bea
{\rho}_{CDW}(E)
=
\frac{E}{2\pi^2} \Big\{&&
\theta(E-q-\Delta) \sqrt{(E-q)^2-\Delta ^2}
\nonumber\\
&+&
\theta(E-q+\Delta)\theta(E+q-\Delta)\sqrt{(E+q)^2-\Delta ^2}
\nonumber\\
&+&
\theta(q-\Delta-E)
\left(\sqrt{(E+q)^2-\Delta^2}-\sqrt{(E-q)^2-\Delta^2}\right)
\Big\}
\,.
\eea

\noindent
In the NJL model, however, the CDW is always less favored than the soliton lattice. 
The main reason to include it in our discussion is its simplicity. 
In particular its density profile is uniform. 
This is most easily seen by applying a global chiral transformation 
of the form $\psi\rightarrow\exp(i\gamma_5\tau_3\,q z_0/2)\psi$
with some constant $z_0$. 
While the density $\langle\psi^\dagger(\x)\psi(\x)\rangle$ is invariant
under this transformation, it leads to a phase shift in $M(x)$
which is equivalent to a translation by $z_0$ in the z-direction.
Hence the density must be uniform. The replacement \eq{eq:napprox} 
is therefore an exact manipulation for the CDW, which will allow us to 
comment on the corresponding approximation for the solitons.

\subsection{Numerical results including the vector-channel interaction}
\label{subsec:numGv}
\noindent
Having set up the formalism for the inclusion of the vector interaction,
we now turn to the numerical part of our investigations.
As in Ref.~\cite{Nickel:2009wj} we fix the parameters 
in the chiral limit by requiring for the pion decay constant 
$f_\pi=88$~MeV and for the constituent mass in the vacuum $M_0=300$~MeV. 
The resulting parameter values are $\Lambda=757.0~{\rm MeV}$ and 
$G_S=6.002/\Lambda^2$.
The value of $G_V$, on the other hand, is treated as a free parameter,
which will be varied in order to study its effect on the phase diagram.
Starting from a color-current interaction and performing a Fierz 
transformation one would get  $G_V/G_S=1/2$, whereas fits to the vector meson 
spectrum typically yield larger values~\cite{Vogl:1991qt,Ebert:1985kz}.
In the following we will therefore vary $G_V$ between zero and 
$G_S$. In particular we will restrict ourselves to repulsive 
vector interactions, similar as in the Walecka model.

\subsubsection{Phase diagrams in the chiral limit}
\label{subsec:phasediag}

\begin{figure}
\begin{center}
\includegraphics[width=.45\textwidth]{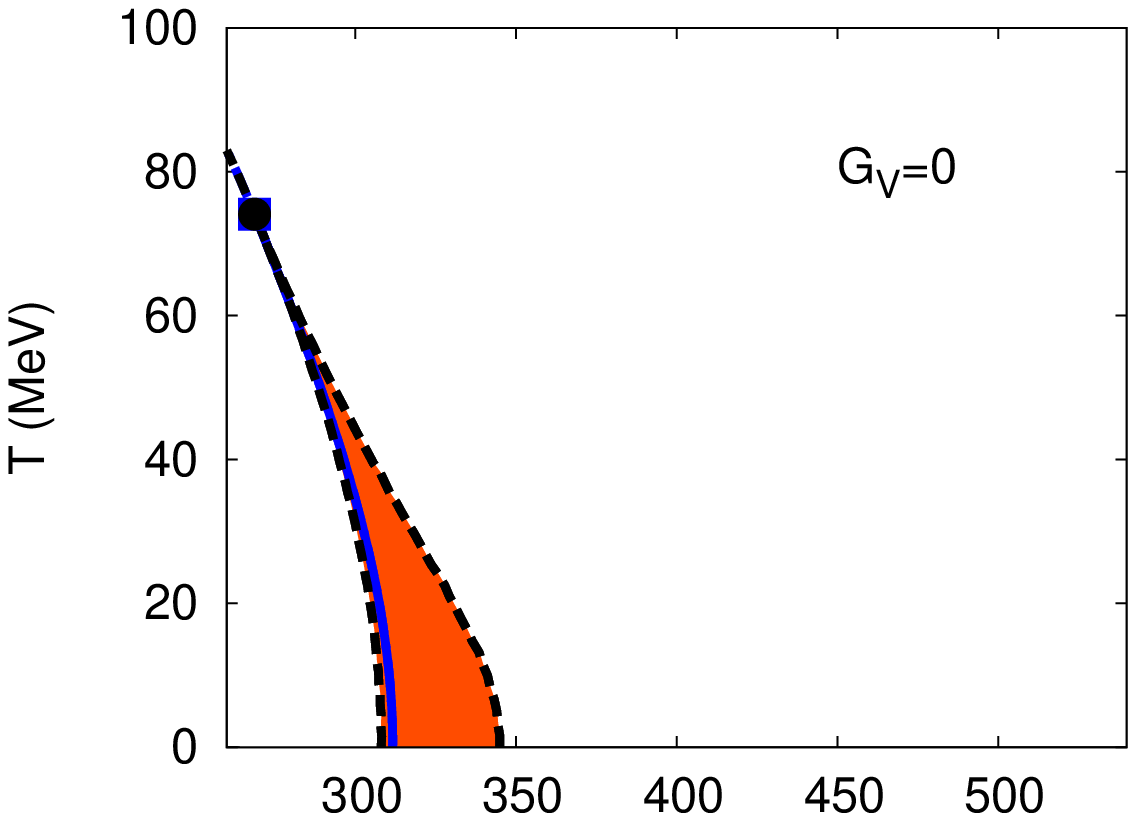}
\includegraphics[width=.45\textwidth]{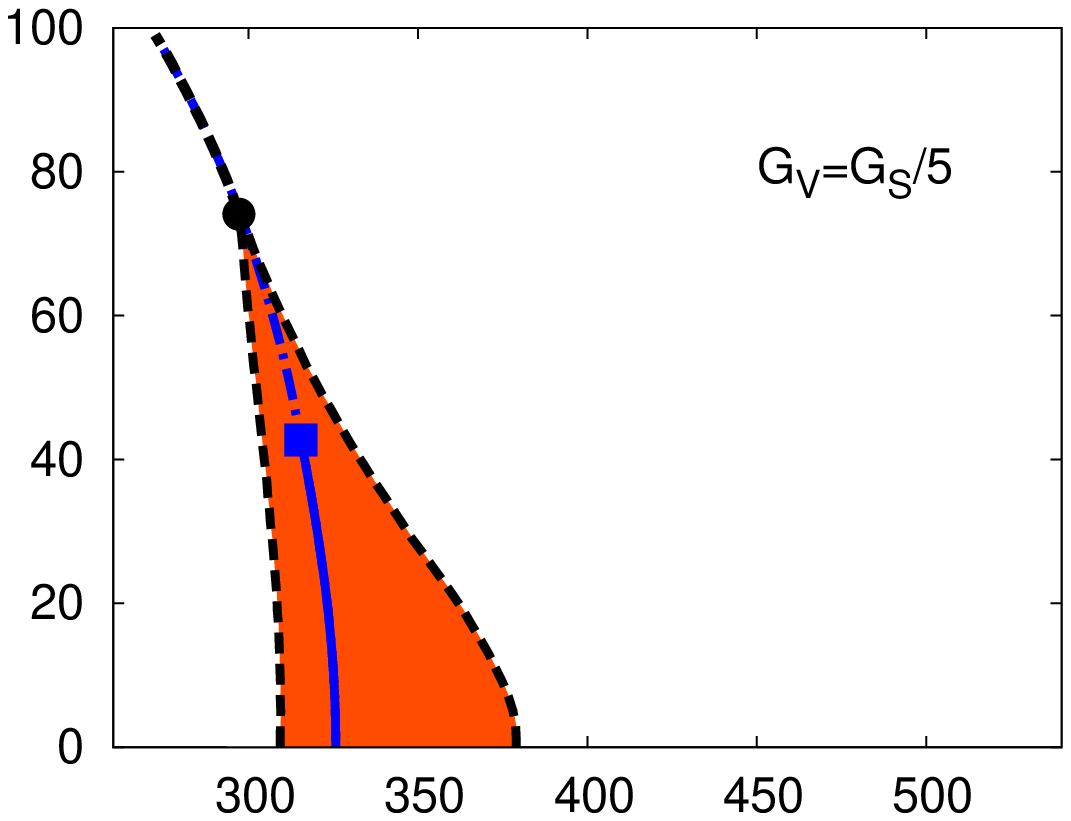}
\includegraphics[width=.45\textwidth]{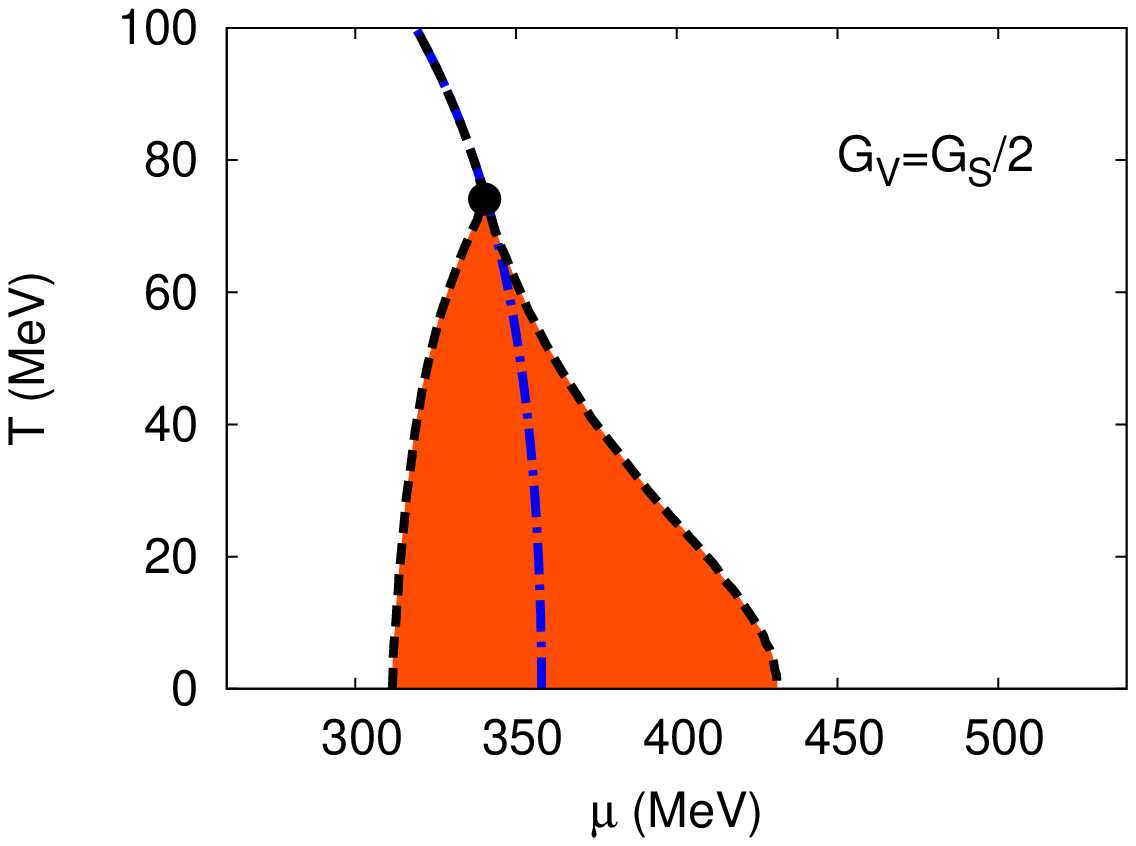}
\includegraphics[width=.45\textwidth]{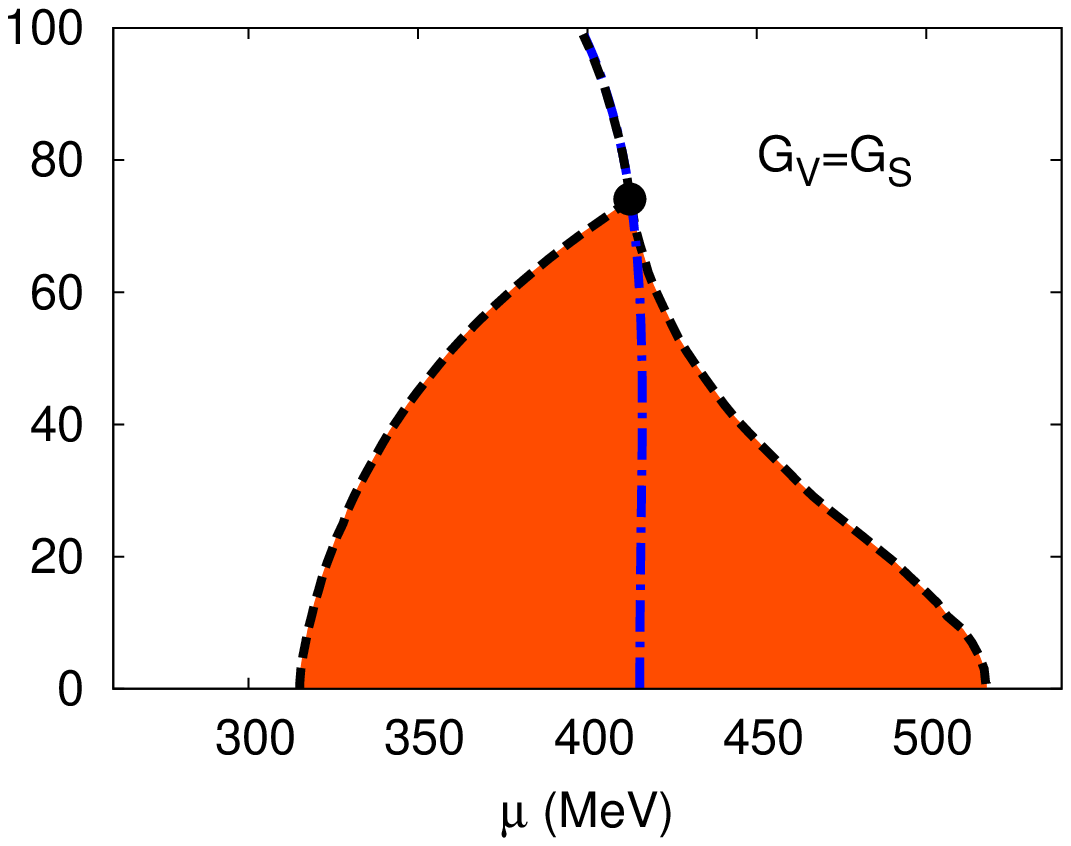}
\caption{The phase diagram in the chiral limit for different values of the 
vector coupling when allowing for the domain-wall soliton lattice. 
The black dashed lines represent the second-order transition lines joining 
at the Lifshitz point  (dot), the shaded region represents the 
inhomogeneous phase. 
The blue solid lines represent the first-order phase 
transition obtained when limiting to homogeneous order parameters, 
which turns to second order (blue dot-dashed lines) at the 
critical point (square).}
\label{fig:pdvector}
\end{center}
\end{figure}

\noindent
Probably the most interesting result of this work is the effect of the 
vector-channel interaction on the transition lines in the phase diagram.
In Fig.~\ref{fig:pdvector} we present the $\mu-T$ phase diagrams for various 
values of $G_V$, focusing on the region where the domain-wall 
soliton lattice is preferred.
The calculations have been performed in the chiral limit. 
For comparison we have also indicated the transition lines one 
obtains when the analysis is limited to homogeneous phases. 
In this case there is a critical point  for small values of $G_V$ (blue square) 
below which the transition from the broken to the restored phase is first 
order (blue solid line).
Upon increasing $G_V$ the CP is shifted to smaller temperatures 
(and higher chemical potentials) and eventually hits the zero temperature 
axis, so that the first-order phase transition is absent for larger values
of $G_V$.
This behavior is well known~\cite{Buballa:1996tm,Kitazawa:2002bc}
and has recently attracted new interest in the discussion of the critical 
surface \cite{Fukushima:2008,Fukushima:2008b}.

\noindent
However, the picture changes considerably, when we allow for inhomogeneous
solutions. We then always find a regime where the domain-wall 
solitons are preferred (shaded region), so that we can distinguish three 
different phases: the homogeneous broken phase, the restored phase and the 
inhomogeneous phase. 
The corresponding three phase boundaries are all of second order.
Their conjunction defines a Lifshitz point (dot), above which the 
phase boundary coincides with that found when limiting to homogeneous phases.
Moreover, for $G_V=0$  the LP precisely agrees with the CP of the purely
homogeneous analysis~\cite{Nickel:2009ke,Nickel:2009wj}.
It turns out, however, that this is no longer true for $G_V>0$:
Whereas with increasing vector coupling the CP moves downwards in 
temperature and eventually disappears from the phase diagram, we 
observe that the Lifshitz point is only shifted in the
$\mu$-direction, while remaining at the same temperature.
Consequently, unlike the first-order boundary in the purely homogeneous
case, the existence of the inhomogeneous phase is not inhibited by the
vector interaction.
At vanishing temperature the transition from the homogeneous broken to the 
inhomogeneous phase is only slightly varying with $G_V$, whereas the 
transition from the inhomogeneous to the restored phase significantly shifts, 
thus enhancing the domain where inhomogeneous phases are favored. 

\noindent
The observed $G_V$-dependence of the phase diagram can easily be understood 
when we recall from the end of section~\ref{subsec:model} that the stationary 
condition (\ref{eq:stationary}) depends on $G_V$ only through $\tilde{\mu}$.
Consequently its solutions $M(\x)$ at given $(\tilde{\mu}$, $T)$ are
independent of $G_V$ and thus equal to the solutions at $G_V=0$ where
$\tilde\mu = \mu$. 
At $G_V > 0$, we can then translate these solutions into solutions
at shifted values of $\mu$, which are obtained from \eq{eq:gapeqmueff}. 
For a given mass function this mapping is unique since the average density 
$\nave$, which enters  \eq{eq:gapeqmueff}, also depends on $G_V$ only through 
$\tilde{\mu}$, see \eq{eq:n}.

\noindent
The consequences of this mapping can be elaborated further for the phase transition lines.
We first consider the case of second-order phase transitions as found in our 
calculation. When approaching the second-order transition from any direction, 
for each extremum in the thermodynamic potential that 
is relevant for the transition the density approaches the same value and 
consequently all extrema are mapped to the same value of $\mu$.
Since there is only one extremum left on one side of the phase transition, 
we only have to make sure that the additional extremum on the other side 
cannot be mapped beyond the phase transition line, thus generating a spinodal 
region.
For the transitions into the restored phase this is guaranteed by the fact 
that the restored solution has the highest possible density at given 
$\tilde{\mu}$ and that the density increases with $\tilde{\mu}$. Therefore no 
other solution can be mapped beyond the transition line for $G_V>0$.
Similarly the homogeneous broken solution has the lowest density at given 
$\tilde{\mu}$ and 
therefore no inhomogeneous solution can be mapped below the transition line 
when going from the inhomogeneous into the homogeneous broken phase.
Consequently all second-order transition lines for $G_V=0$ are mapped onto 
second-order transition lines for $G_V>0$. In particular, this explains why 
the Lifshitz point stays at the same temperature, as the mapping leaves $T$ 
untouched.

\begin{figure}
\begin{center}
\includegraphics[width=.6\textwidth]{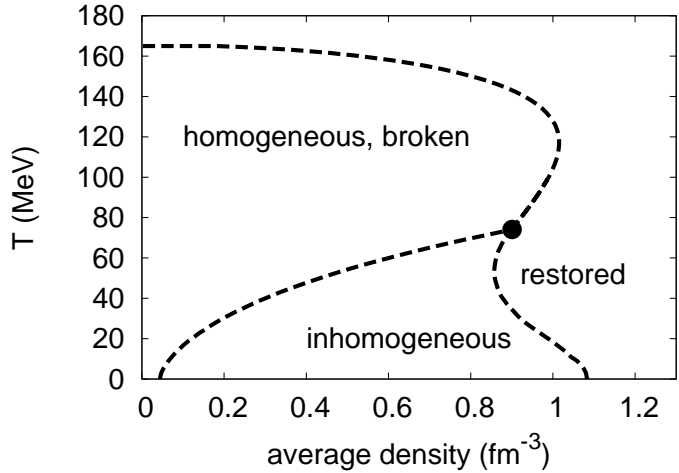}
\caption{The phase diagram in the $\nave - T$ plane. 
The transition lines do not depend on $G_V$ for $G_V>0$.}
\label{fig:Tvsn}
\end{center}
\end{figure}

\noindent
From the arguments given above, it follows immediately that the phase diagram 
in the $\tilde{\mu}$-$T$ plane is independent of $G_V$. 
Moreover, since in this case $\nave(T,\tilde\mu)$ is uniquely given
by \eq{eq:n} and therefore also independent of $G_V$,
this means that the $\nave$-$T$ phase diagram is independent of
$G_V$ as well. This diagram is presented in Fig.~\ref{fig:Tvsn}.

\noindent
A slight complication of this picture arises if there is a 
first-order phase transition at $G_V=0$, which occurs when limiting to 
homogeneous phases.
In this case we have a spinodal region enclosing the first-order phase 
transition line, where the thermodynamic potential has several extrema 
at the same value of $\mu$: two local minima and one local maximum.
As before, this means that at $G_V>0$ we have several solutions at the
same value of $\tilde\mu$. 
However, since the local extrema correspond to different masses and 
therefore to different densities, they will now be mapped onto 
different values of $\mu$ by \eq{eq:gapeqmueff}: More precisely,
solutions with lower masses will be shifted to higher values of
$\mu$ than solutions with higher masses. 
As a consequence, the spinodal region shrinks with increasing $G_V$,
i.e., at fixed temperature the first-order transition gets weakened
and eventually becomes second order. 
This explains why the CP moves to lower temperatures and finally disappears 
from the phase diagram. 
(See also Ref.~\cite{Fukushima:2008b} for a detailed discussion of this
effect.)

\noindent
Finally, we would like to comment on the slopes of the three phase 
transition lines in the LP. For $G_V=0$ all of them are equal, i.e.,
the phase boundaries are tangential in the LP. In the $\tmu-T$ plane, 
this remains of course true for $G_V>0$.
However, when we employ \eq{eq:gapeqmueff} to map $\tmu$ onto $\mu$
we have to keep in mind that the average density $\nave$ at given $T$ and 
$\tmu$ depends on the constituent quark mass $M$.
The latter vanishes identically at the two phase boundaries to the 
restored phase, but is nonzero at the boundary between the homogeneous 
broken and the inhomogeneous phase, approaching zero only towards the LP.
Depending on the corresponding critical exponent, this can affect the
slope of this boundary in such a way that it meets the two others
with a nonvanishing angle at $G_V>0$.
As one can see in Fig.~\ref{fig:pdvector}, this is indeed the case.

\subsubsection{Density profiles}

\noindent
The results presented above are based on the assumption that 
in the thermodynamic potential the density $n(\x)$ can be 
approximated by its spatial average $\nave$, see \eq{eq:napprox}.
Of course this approximation might appear questionable in an 
inhomogeneous phase. In order to get some insight into this issue,
we now want to discuss the density profile $n(z)$ of the solitons.
To that end we restrict ourselves to the case $G_V=0$, where 
\eq{eq:napprox} does not enter into the derivation.
To the extent that \eq{eq:napprox} is a good approximation, the results
can then be translated to $G_V>0$ by the mapping \eq{eq:gapeqmueff}.
For simplicity, we will also limit ourselves to the chiral limit,
although this is not essential.

\noindent
In mean-field the density profile is given by the expectation value
\bea
n(\x)
&=&
\langle \psi^\dagger(\x)\psi(\x) \rangle
\nonumber\\
&=&
\frac{1}{2V}
\sum_{E_i}
\psi^\dagger_{E_i}(\x)\psi_{E_i}^{\phantom{\dagger}}(\x)
\left(n_+(E_i) - n_-(E_i) \right) \,,
\label{eq:density0}
\eea
where the $\psi_{E_i}(\x)$ are eigenfunctions of the Hamiltonian $H_0$
(see \eq{eq:hamiltonian0}) for the eigenvalues $E_i$ and $n_\pm(E)$ are 
the Fermi occupation numbers defined below \eq{eq:n}.

\noindent
For one-dimensional modulations the energy spectrum is highly degenerate 
and we can label the eigenvalues as $E=\sqrt{\lambda^2+\p_\perp^2}$, 
where $\p_\perp$ is the conserved momentum of the quasi-particle perpendicular 
to the modulation and $\lambda$ is the eigenvalue of the Hamiltonian at 
$\p_\perp=0$. The latter reduces to the Gross-Neveau Hamiltonian and we refer 
to Ref.~\cite{Nickel:2009wj} for details. Since 
$\psi^\dagger_{E}(z)\psi_{E}^{\phantom{\dagger}}(z)$ in this case only 
depends on $\lambda$, we arrive at
\bea
n(z)
&=&
\frac{2N_c}{V}
\int d\lambda\, \rho_1(\lambda)
\int\frac{dp_\perp}{(2\pi)^{d_\perp}}
\psi^\dagger_{\lambda}(z)
\psi^{\phantom{\dagger}}_{\lambda}(z) \left(n_+(E) - n_-(E) \right) \,,
\label{eq:density}
\eea
where $\rho_1(\lambda)$ is the spectral density of the one-dimensional 
GN model~\cite{Schnetz:2004vr,Schnetz:2005ih,Thies:2006ti,Basar:2009fg}
and
\bea
\psi^\dagger_{\lambda}(z)\psi_{\lambda}(z)
&=&
\frac{(\lambda/\Delta)^2+\frac{1}{2}((M(z)/\Delta)^2 + \nu -2)}
{(\lambda/\Delta)^2 -\E(\nu)/\K(\nu)}
\,.
\eea
Similar to the determination of the effective density of states, \eq{eq:rho} 
(see  Ref.~\cite{Nickel:2009wj} for details), it is then a tedious, but 
straightforward exercise to cast the expression for the density profile into 
the form
\bea
n_{\rm soliton}(z)
&=&
2N_c
\int_0^\infty  dE\, \rho_{D,{\rm soliton}}(E,z) \left(n_+(E) - n_-(E)\right)
\,,
\eea
where the density matrix element $\rho_{D,{\rm soliton}}(E,z)$ can be related 
to $\rho_{\rm soliton}(E)$, \eq{eq:rho}, upon the replacement
\bea
\rho_{D,{\rm soliton}}(E,z)
&=&
\rho_{\rm soliton}(E) \Big\vert_{
\frac{ {\mathbf E}(\nu )}{{\mathbf K}(\nu )} \rightarrow 
-\frac{1}{2} \left( \left(\frac{M(z)}{\Delta}\right)^2 + \nu -2 \right)
}
\,.
\eea

\begin{figure}
\begin{center}
\includegraphics[width=.24\textwidth]{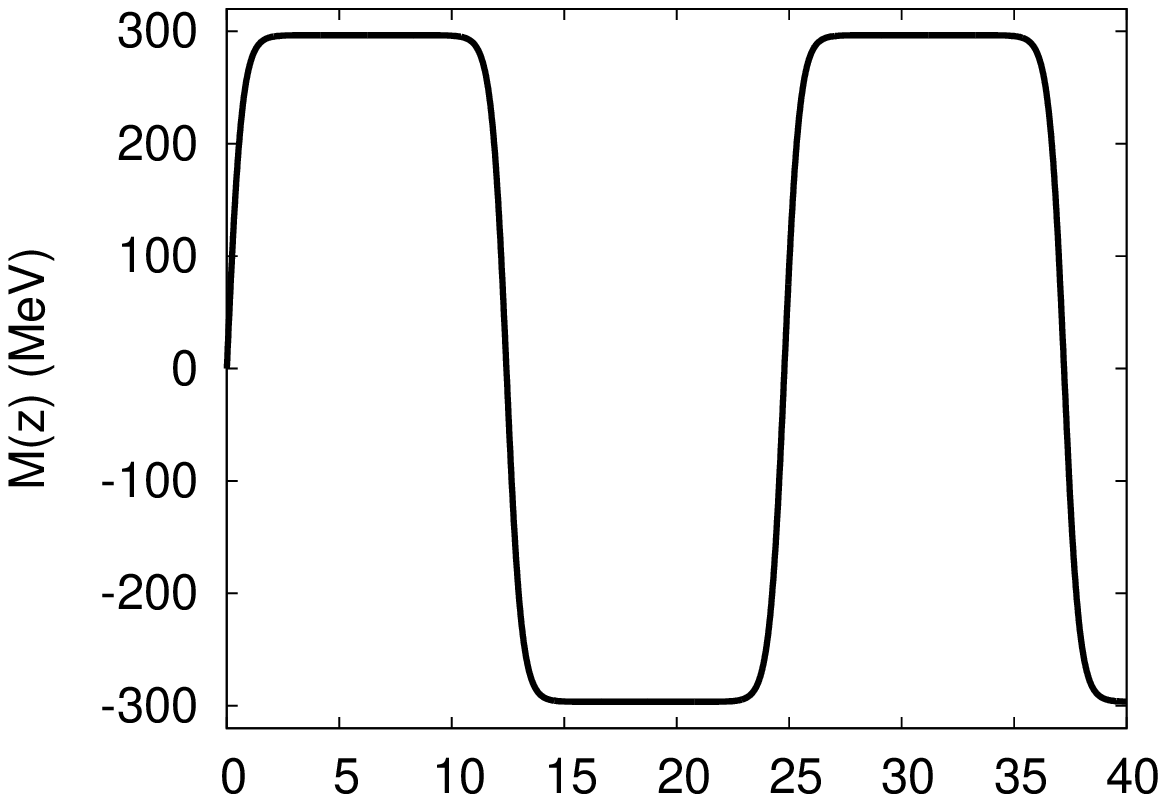}
\includegraphics[width=.24\textwidth]{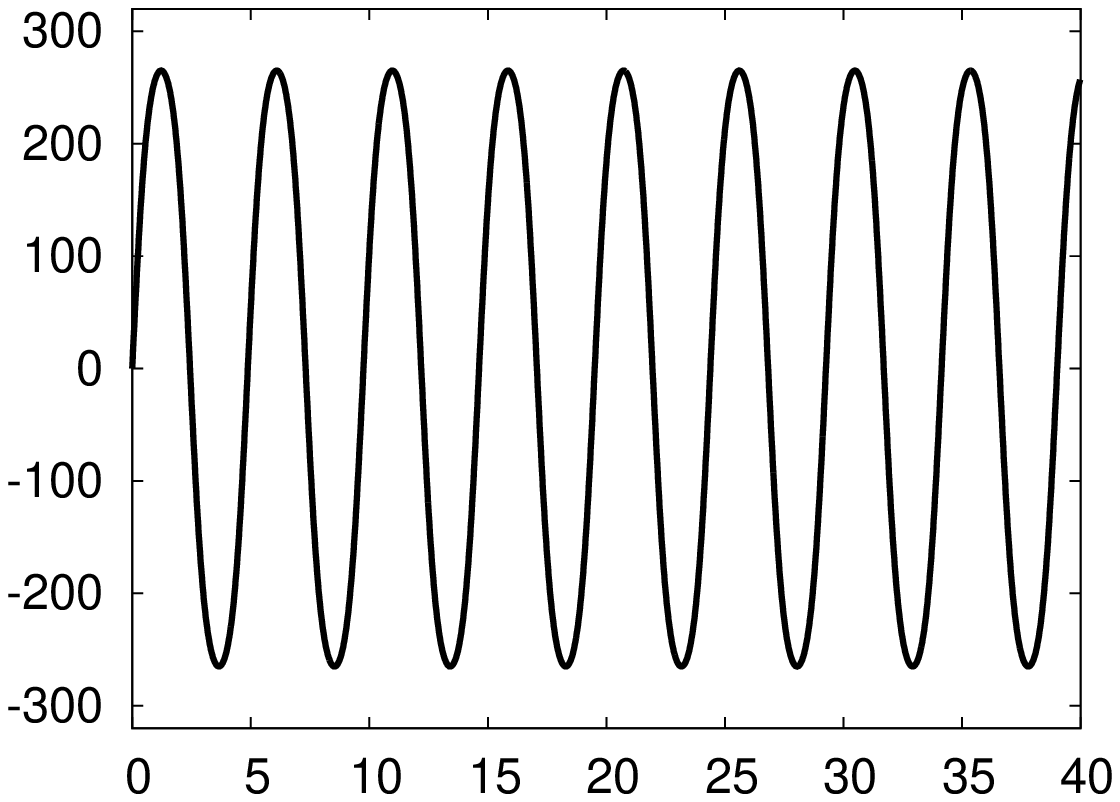}
\includegraphics[width=.24\textwidth]{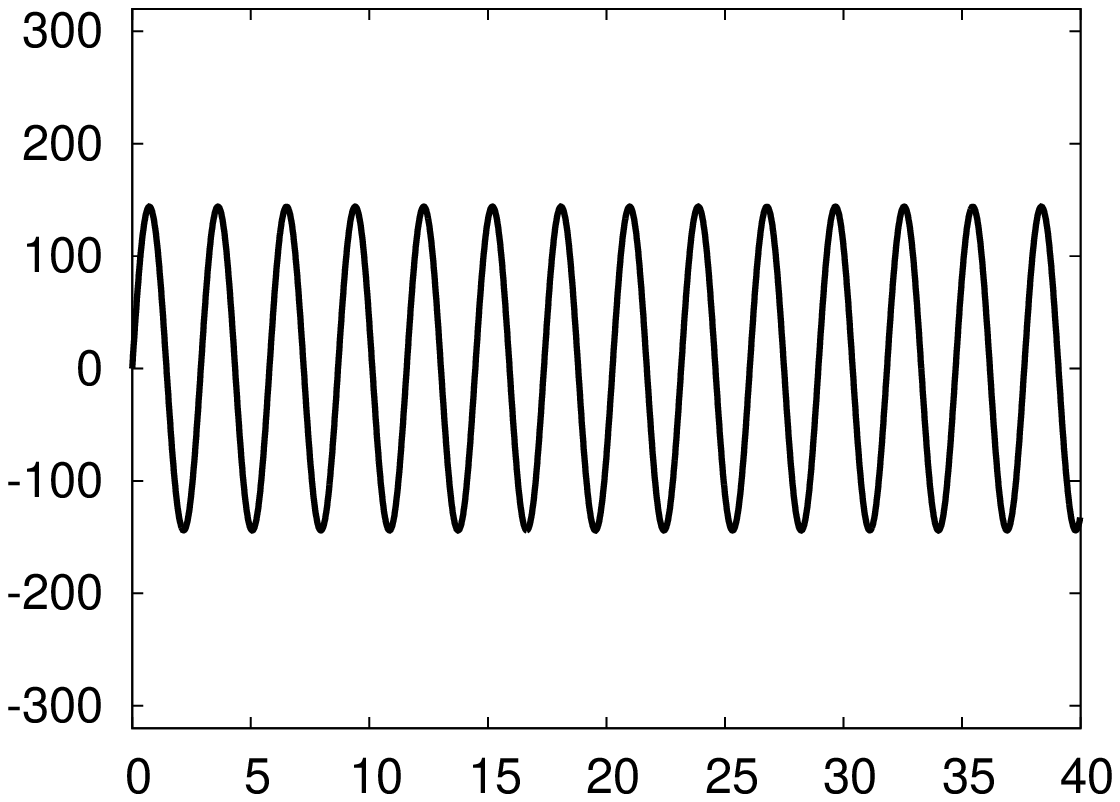}
\includegraphics[width=.24\textwidth]{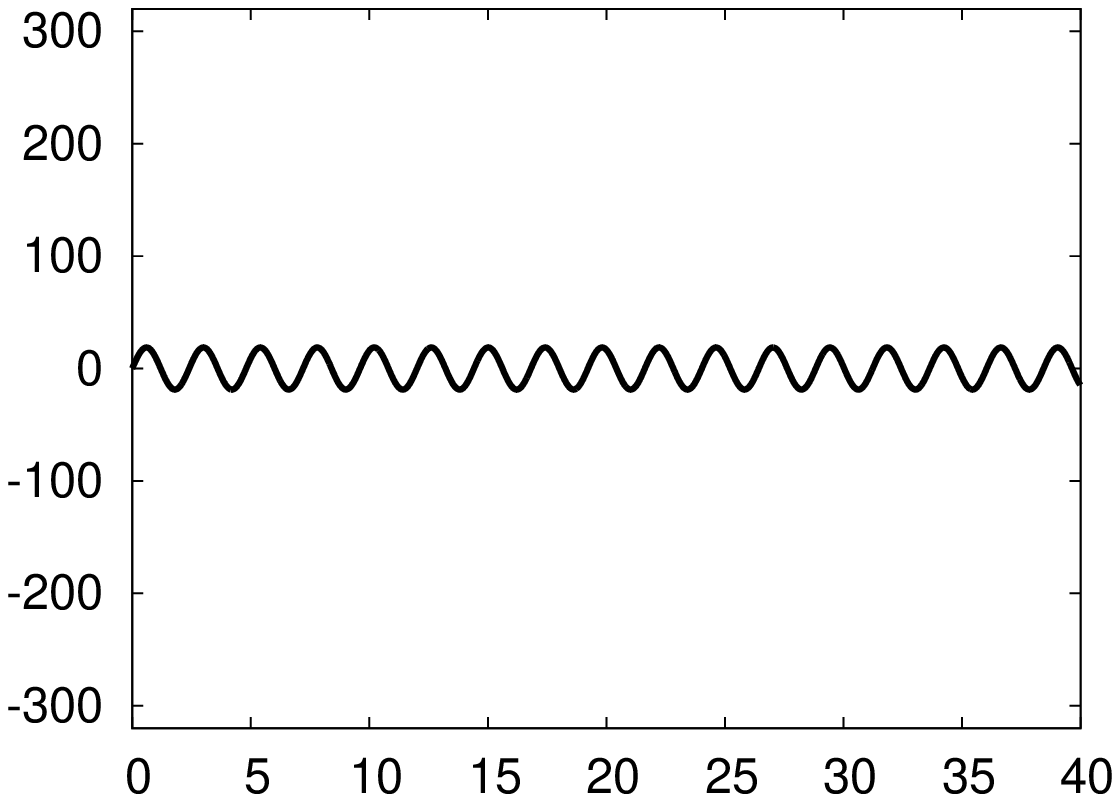}

\includegraphics[width=.24\textwidth]{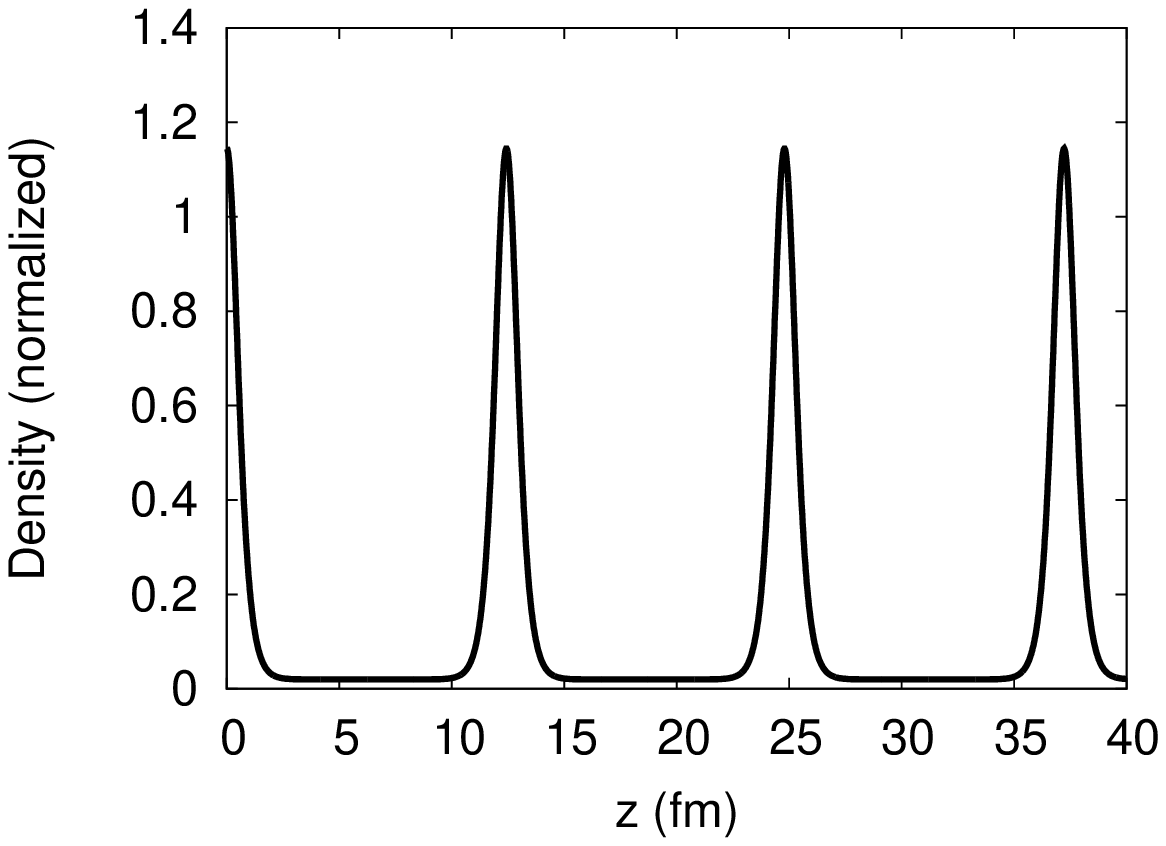}
\includegraphics[width=.24\textwidth]{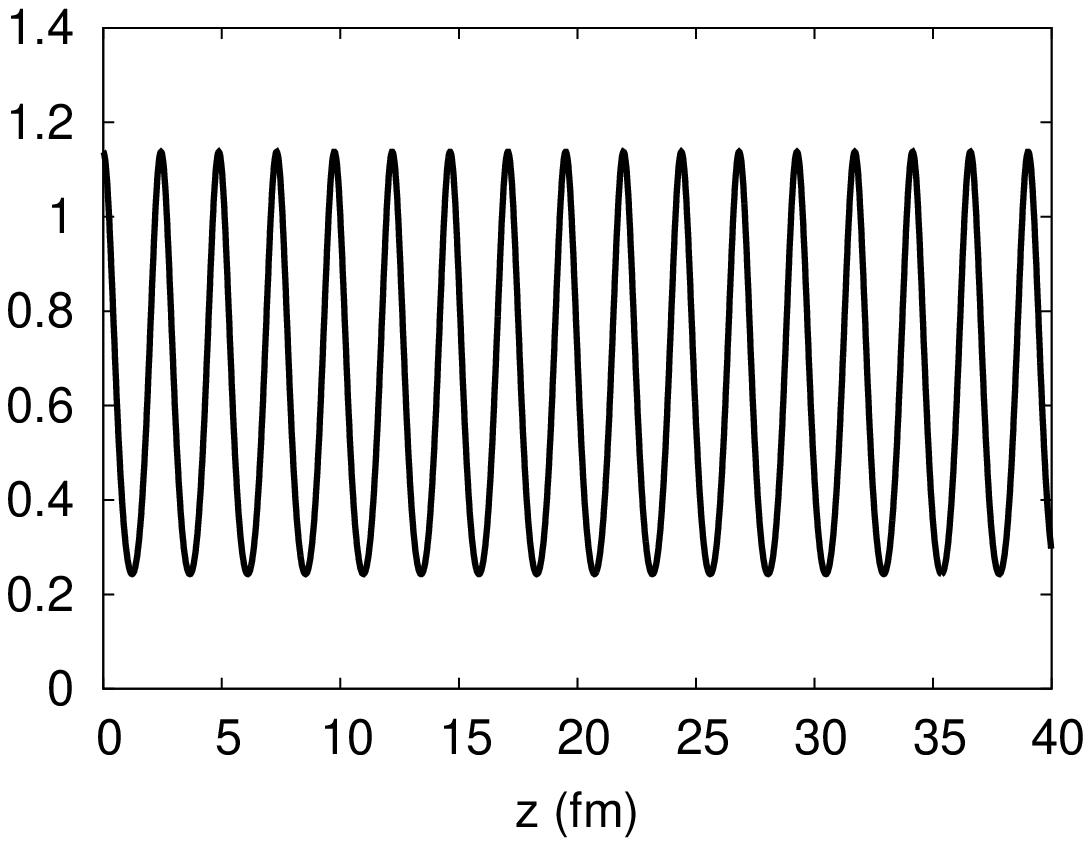}
\includegraphics[width=.24\textwidth]{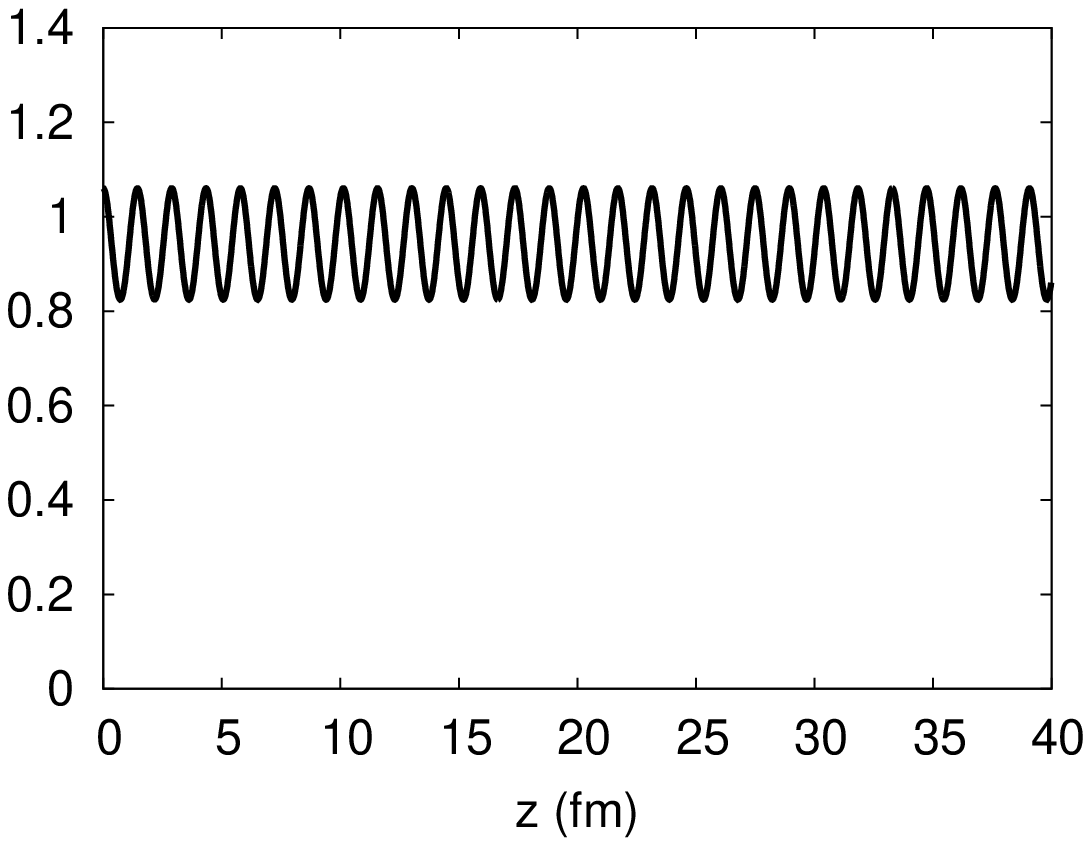}
\includegraphics[width=.24\textwidth]{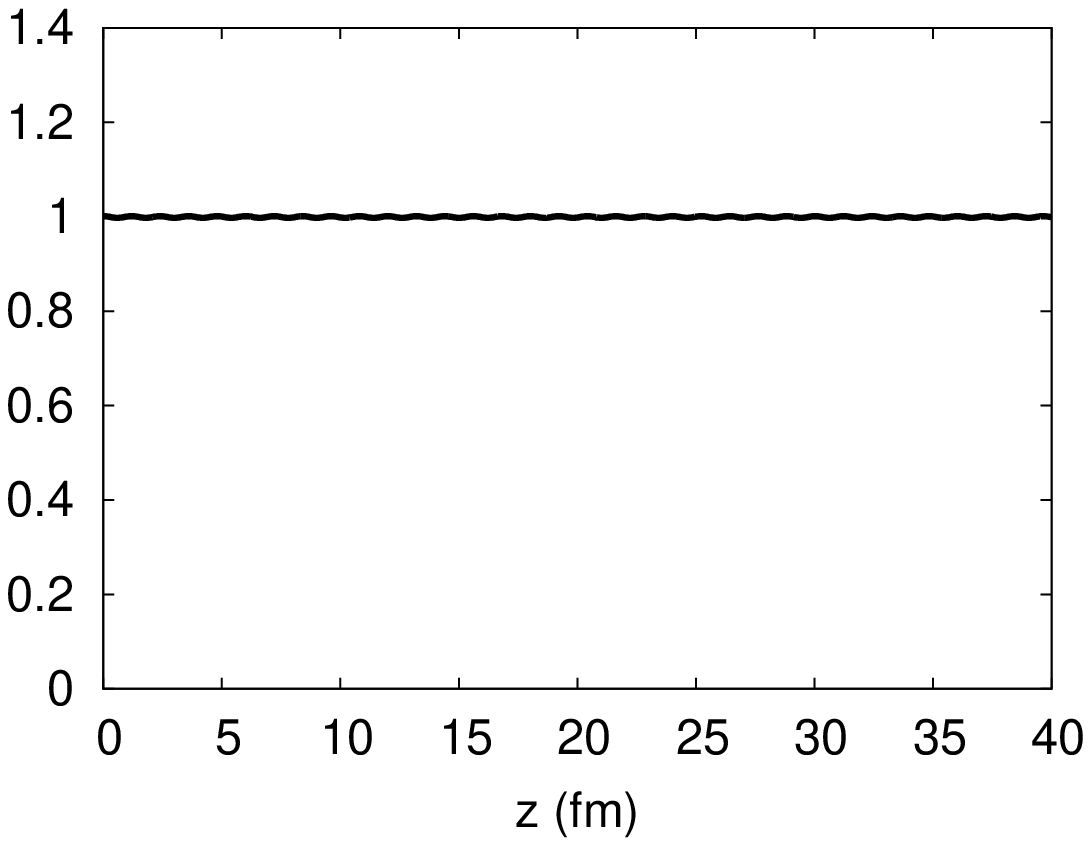}
\caption{
Spatially dependent constituent mass function $M(z)$ (upper row) 
and corresponding density profile (lower row) 
at $G_V=0$ and $m=0$ for $T=0$ and, from left to right, 
$\mu=307.5$, $309$, $325$, and $345$~MeV. 
The density is normalized to the density in the restored phase,
$n_\mathit{rest} = \frac{2N_c}{3\pi^2}\mu^3$.
}
\label{fig:densvsmass}
\end{center}
\end{figure}

\noindent
The resulting density profiles at $T=0$ and four different chemical 
potentials are shown in the lower part of Fig.~\ref{fig:densvsmass}. 
Comparing them with the corresponding mass functions $M(z)$, which are 
displayed in the upper part of the figure, we find that the density
distributions follow closely the positions of the solitons, i.e., of the 
zero-crossings of the mass functions. 
In a bag-model like picture, this can be interpreted as the quarks being 
squeezed by the bag pressure of the domains with broken chiral symmetry 
into the regions of space where chiral symmetry is almost restored. 
From a topological point of view, these quarks are related to zero energy modes localized on the domain-wall.

\noindent
These features are seen most clearly at $\mu = 307.5$~MeV, which is
just above the phase boundary from the homogeneous broken phase. 
Here the solitons are well separated, leading to strongly localized
density peaks as functions of $z$.
In this regime the assuption of a homogeneous density in order to obtain 
the solutions for $G_V>0$ is certainly poor. 
However, when $\mu$ is increased the solitons quickly start to overlap
and the density profiles become more and more washed out. 
At the second-order transition to the restored phase, the order 
parameter melts and the density profile smoothly approaches the uniform 
density of the restored phase. This is illustrated by the example of 
$\mu=345$~MeV, which is close to the phase boundary.
Note, however, that already at $\mu=325$~MeV the density variations are 
relatively small.

\noindent
The $T=0$ results shown in Fig.~\ref{fig:densvsmass} are the most
extreme cases. With increasing temperature the mass functions melt
and the density profiles become  washed out by thermal effects.
The assumption of a uniform density profile is therefore
most questionable at the transition from the homogeneous broken to the 
inhomogeneous phase at low temperatures. In fact, at $G_V>0$,
a local density approximation would result in a lower value of 
$\tilde\mu$ within the solitons.
This reflects the repulsive nature of the vector interaction,
which disfavors localized density peaks.
We therefore expect that the formation of well separated solitons,
as we find near the boundary to the homogeneous broken phase at $T=0$,
eventually becomes inhibited by the vector interaction.
Related to this, the second-order phase transition from the homogeneous
broken to the inhomogeneous phase may partially turn into a first-order 
transition at $G_V > 0$.

\noindent
On the other hand our assumption of a uniform density becomes gradually
better with increasing $\mu$ or $T$ and is fully justified at the 
phase transition line to the restored phase. In particular the phase
boundary itself and the Lifshitz point are not affected by the 
approximation. This will be shown more rigorously in Sec.~\ref{sec:GL}
by a Ginzburg-Landau analysis.

\subsubsection{Chiral density wave}

\begin{figure}
\includegraphics[width=.47\textwidth]{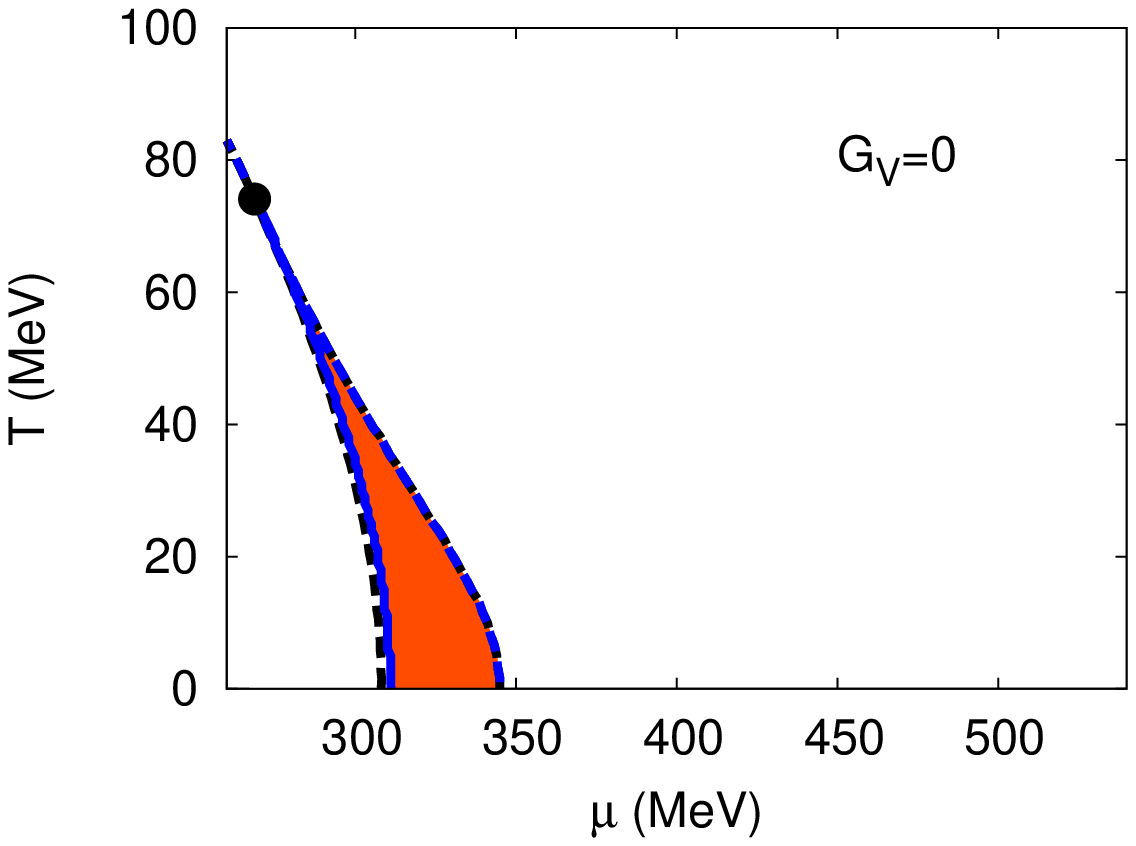}
\includegraphics[width=.47\textwidth]{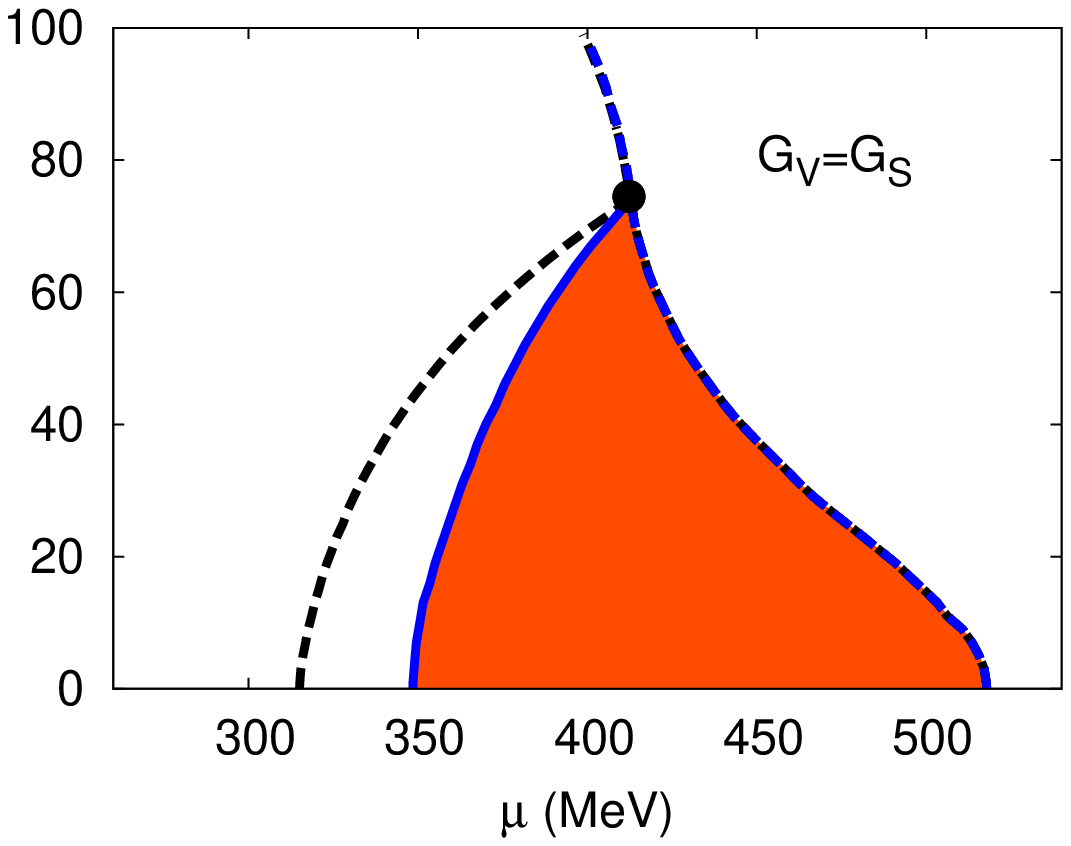}
\caption{Phase diagrams in the $\mu-T$ plane at $G_V=0$ (left) and 
$G_V=G_S$ (right). 
The dashed lines have the same meaning as in Fig.~\ref{fig:pdvector}
and represent the second-order phase boundaries between the homogeneous
broken phase, the restored phase and the inhomogeneous phase with 
domain-wall solitons, respectively. 
The shaded area indicates the region where the CDW is
favored compared to homogeneous phases (but not to the solitons).
Here the blue solid line represents the first-order phase boundary from 
the broken homogeneous phase to the CDW, while the phase boundary from the
CDW to the restored phase is second order and coincides with the boundary
from the soliton phase to the restored phase.}
\label{fig:CDW}
\end{figure}

\noindent
Complementary support for the results of Sec.~\ref{subsec:phasediag}
can be obtained from investigations of the chiral spiral, \eq{eq:MCDW}.
Although, at least at $G_V = 0$, this kind of modulation is disfavored 
compared to the domain-wall soliton~\cite{Nickel:2009wj}, a constant 
density profile is no assumption here, but a property of the state. 
We can therefore perform a completely self-consistent mean-field calculation, 
leading to the phase diagrams shown in Fig.~\ref{fig:CDW}.
The calculations have again been performed in the chiral limit and 
for $G_V=0$ (left panel) and $G_V=G_S$ (right panel).
The regions where the CDW is favored against the homogeneous broken and
restored phases are indicated by the shaded areas. 
For comparison we have also indicated the boundaries of the regime
where the solitons are favored (dashed lines).  

\noindent
As discussed in section~\ref{sec:GL} below, the transition from the 
chiral spiral to the chirally restored phase is second order and agrees 
exactly with the transition from the soliton lattice to the restored phase.
In particular, this also holds for the Lifshitz point. 
On the other hand, the transition from the homogeneous broken phase to the 
state where the chiral spiral is preferred is first order. For this reason, 
given the arguments of Sec.~\ref{subsec:phasediag}, 
it is directly depending on $G_V$, not only through $\tilde{\mu}$. 
As a result, the phase boundary, which at $G_V = 0$ almost coincides with 
the corresponding phase boundary of the soliton phase, moves away from it
at $G_V=G_S$. However, this effect is rather mild.
Moreover we find that the first-order transition line from the homogeneous broken 
phase to the chiral spiral is much less affected by the vector interaction 
than the second-order transition line from the chiral spiral to the restored 
phase. 
The qualitative behavior of the phase diagram as a function of $G_V$ 
is therefore similar to our results for the soliton lattice.
Since we expect the CDW to be disfavored compared to the 
solitons, we can take the CDW result as a lower limit for the 
area occupied by an inhomogeneous phase. 
This suggests that the effect of assuming a uniform 
density profile for the solitons is not drastic.

\subsubsection{susceptibilites}

\begin{figure}
\begin{center}
\includegraphics[width=.48\textwidth]{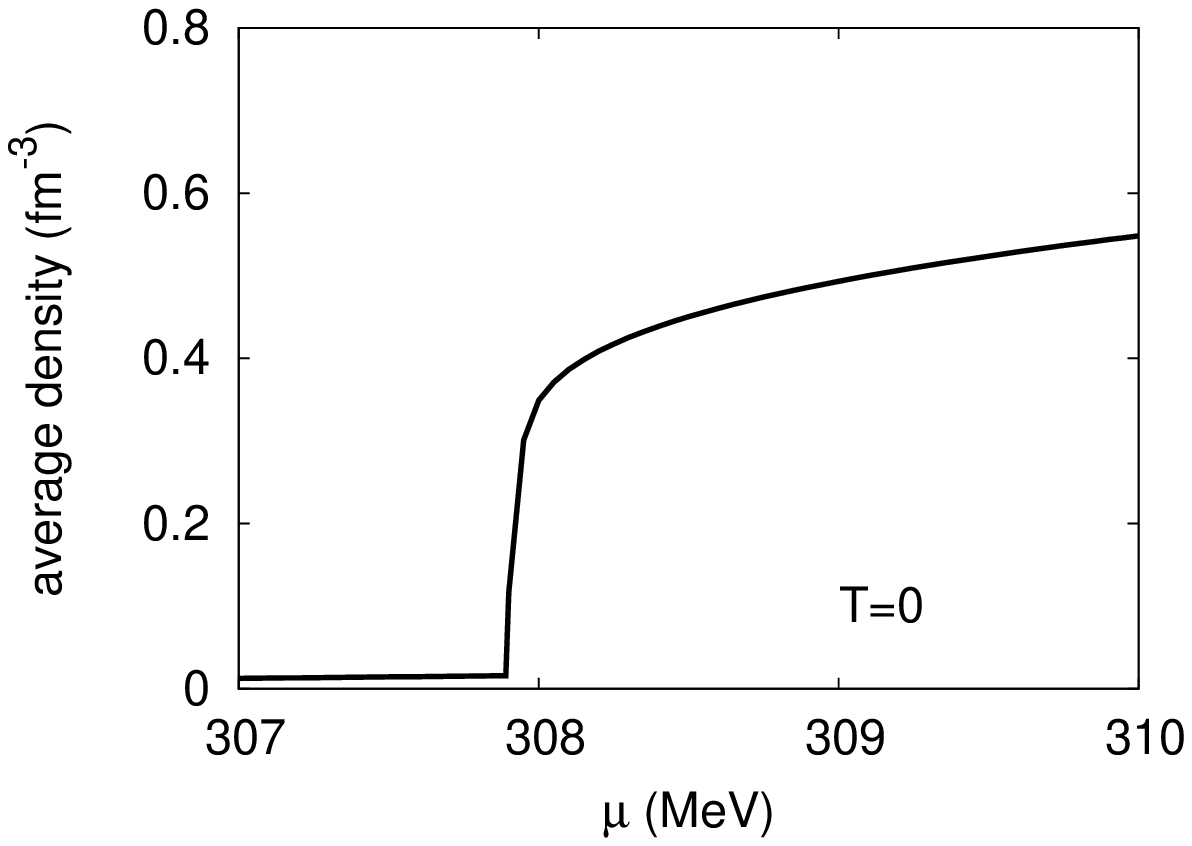}
\includegraphics[width=.48\textwidth]{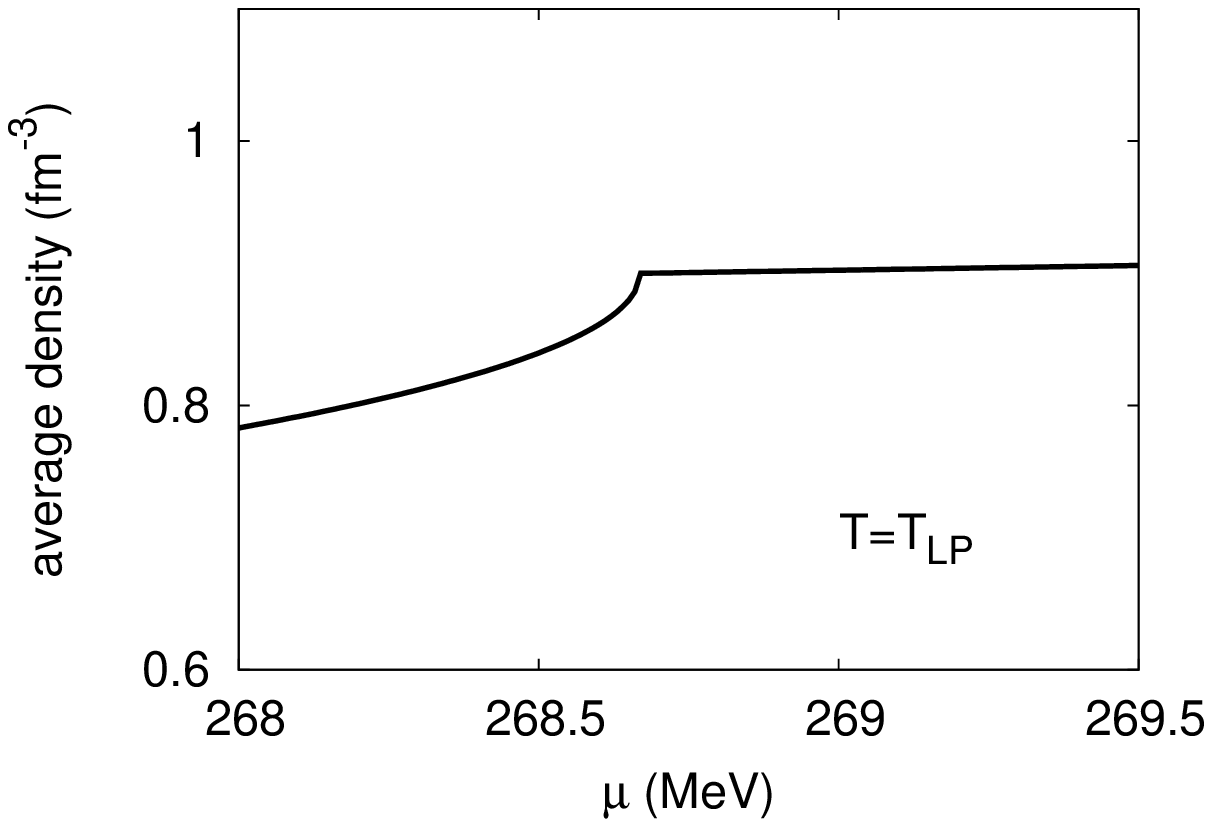}

\includegraphics[width=.48\textwidth]{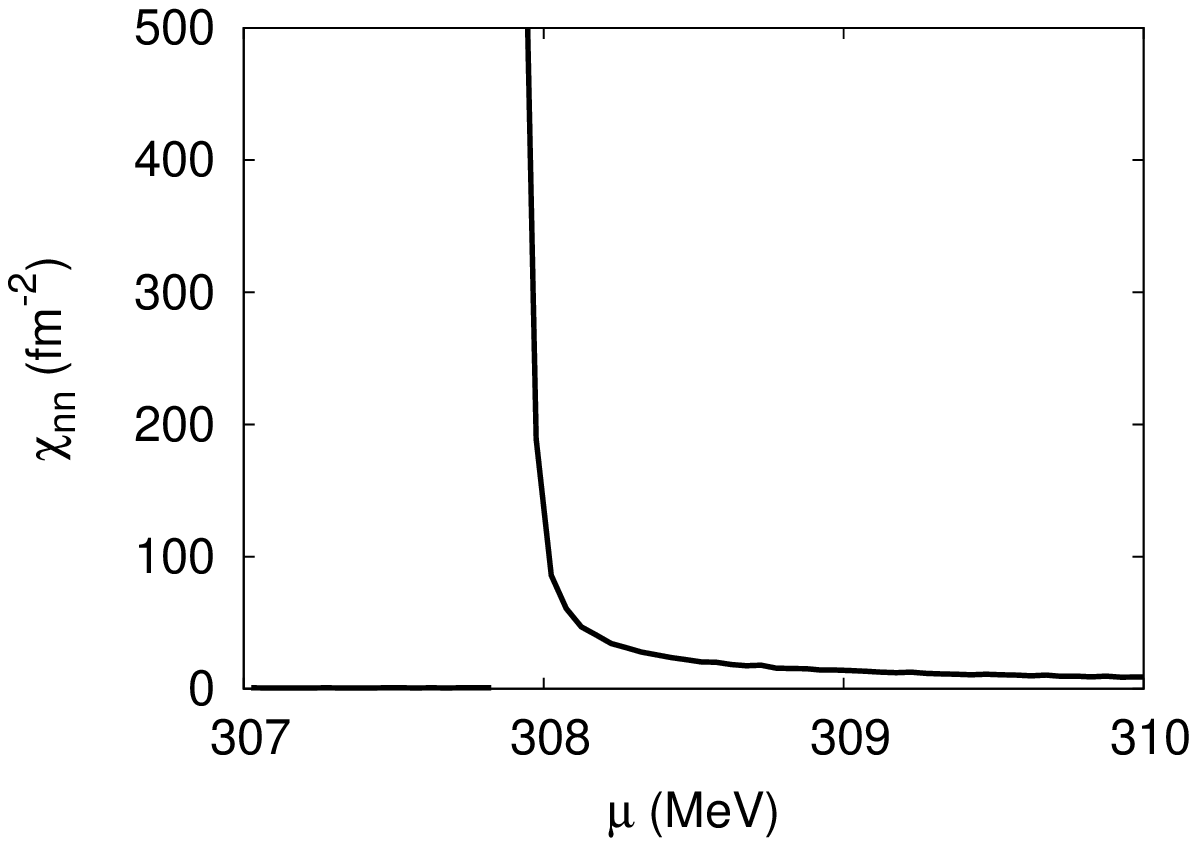}
\includegraphics[width=.48\textwidth]{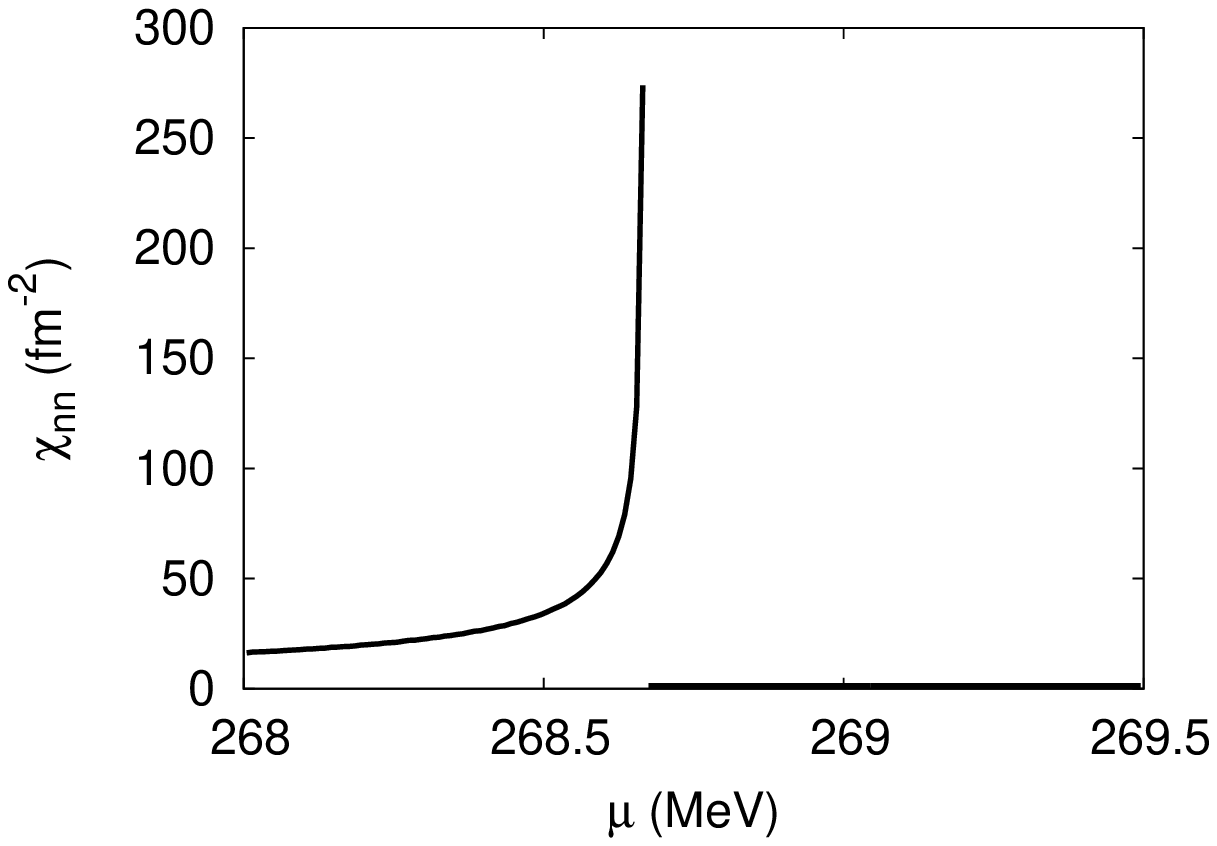}
\caption{
Averaged density (upper panel) and number susceptibility (lower panel) for vanishing temperatures (left) and for the temperature at the LP (right) as a function of chemical potential and for $G_V=0$.
}
\label{fig:denssusc}
\end{center}
\end{figure}

\noindent
The divergence of susceptibilities near a critical point has led to suggestions for how to locate the CP in the QCD phase diagram experimentally ~\cite{Stephanov:1998dy} and therefore attracted significant interest also in model studies.
Since the CP as a cornerstone of the phase diagram is essentially replaced by the LP in our study, we first want to discuss the behavior in its vicinity.
For simplicity we will limit ourselves to the number susceptibility
\begin{align}
\chi_{nn}
\quad=\quad
-\frac{
\partial^2 \Omega
}{
\partial\mu^2
}
\quad=\quad
\frac{\partial \nave}{\partial \mu}
\,,
\end{align}
which corresponds to the change in density when going along a line of constant temperature.
For this reason the behavior of $\chi_{nn}$ near the LP is in fact determined by the behavior when going from the homogeneous broken to the restored phase and not by the inhomogeneous phase.
Consequently, since the LP coincides with the CP for $G_V=0$, we find the same divergent behavior as when limiting to homogenous phases for $G_V=0$. For completeness, this is illustrated on the right hand side of Fig.~\ref{fig:denssusc}, showing a $1/\sqrt{\mu_{\rm cr}-\mu}$-like singulartity when approaching the LP from the left.
However, for $G_V>0$ the behavior is qualitatively altered: The would-be CP when limiting to homogeneous phases is hidden inside the domain of a inhomogeneous phase and the LP does not correspond to the endpoint of the second-order phase transition line when limiting to homogeneous phases.
For this reason the number susceptibility does not diverge near the LP for $G_V>0$.
This can easily be understood in the context of a Ginzburg-Landau expansion as introduced in subsection~\ref{sec:GL} and is related to the fact that $c_{4,a}\neq0$ for $G_V>0$ at the LP.

\begin{figure}
\begin{center}
\includegraphics[width=.48\textwidth]{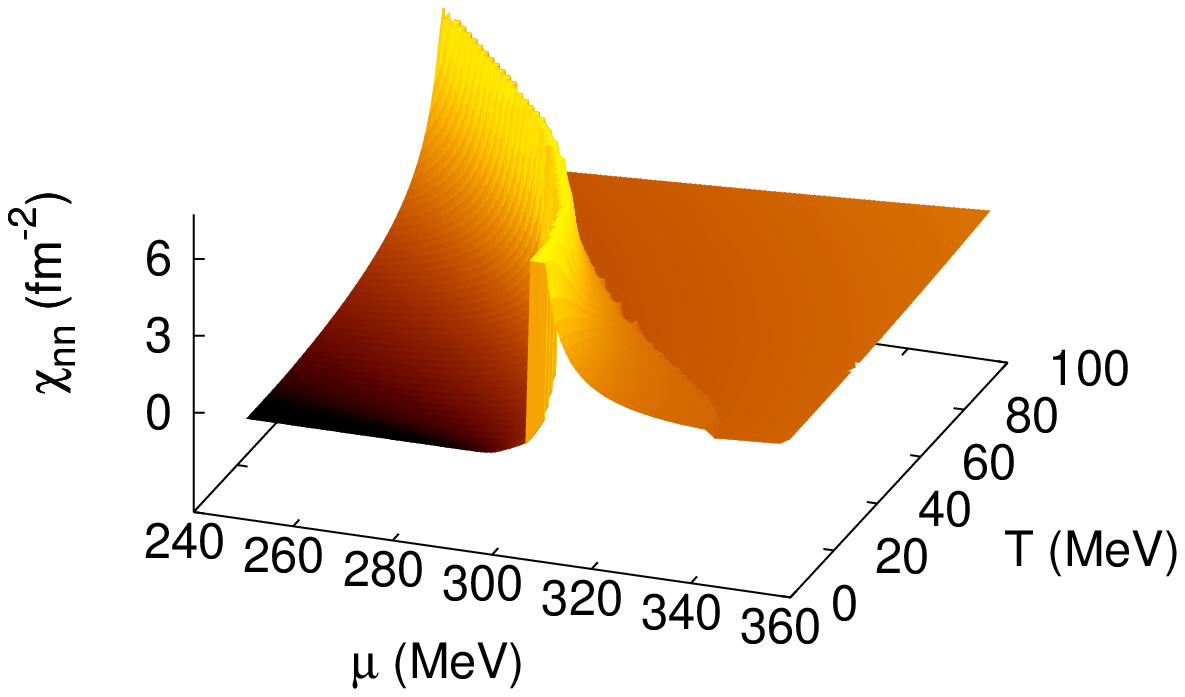}
\includegraphics[width=.48\textwidth]{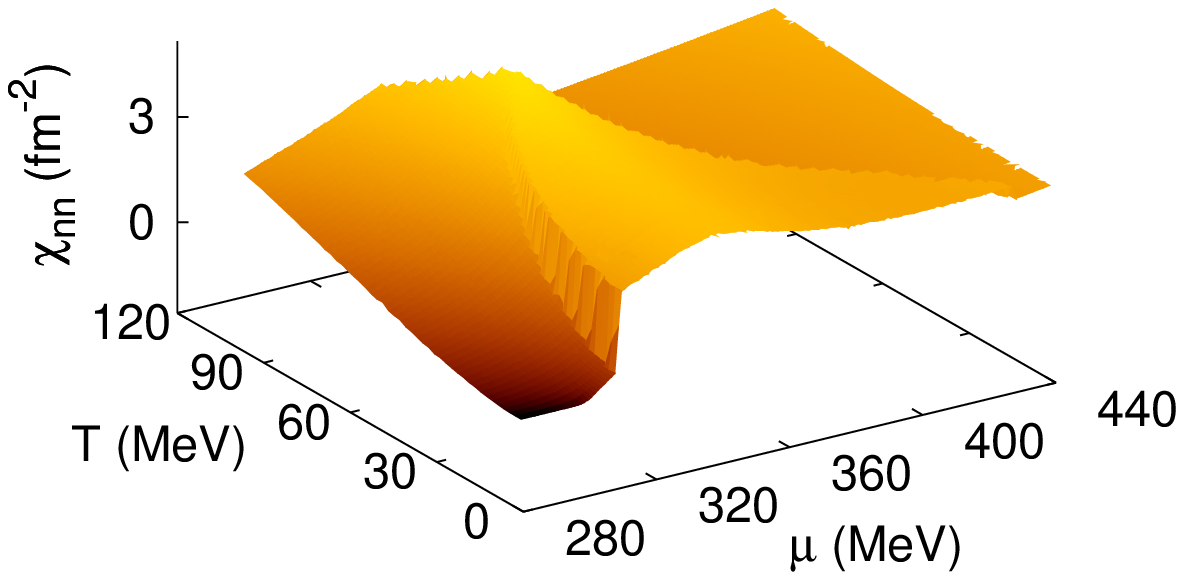}
\caption{
Number susceptibility in the $\mu-T$ plane at $G_V=0$ (left) and $G_V=G_S/2$ (right). For $G_V=0$ the number susceptibility diverges at the transition from broken to inhomogeneous phase.
}
\label{fig:susc3d}
\end{center}
\end{figure}

\noindent
Focusing on $G_V=0$ first and investigating the number susceptibility numerically in the regime where soliton lattices are energetically preferred, we find the results shown in Figs.~\ref{fig:denssusc},~\ref{fig:susc3d}.
Since all transition are second order, the averaged number density changes continuously when varying temperature and chemical potential. At the transition from the homogeneous broken to the inhomogeneous phase the change is however very rapid as discussed in the following and $\chi_{nn}$ diverges. This is most prominent at $T=0$ and decreases when going towards the LP.

\noindent
In order to get an intuition for the qualitative behavior of $\chi_{nn}$ we consider the GN model and focus on the density as a function of chemical potential $n_{GN}(\mu)$ at vanishing temperatures. As can be extracted from Refs.~\cite{Schnetz:2004vr,Thies:2006ti} it is given by
\bea
n_{GN}
&=&
\frac{\pi^2}{2\nu \K(\nu)} M_0
\,,
\eea
where $M_0$ as the fermion mass in the vacuum sets the scale and the elliptic modulus as a function of chemical potential is given through the implicit relation $\pi \sqrt{\nu} \mu=2\E(\nu)M_0$. At the transition from homogeneous to inhomogeneous phase at $\mu_{cr}=\frac{2}{\pi}M_0$ the number density then behaves like
\bea
n_{GN}
&\simeq&
-\frac{\pi^2 M_0}{\ln(\mu/\mu_{cr}-1)}
\,,
\label{eq:nGN}
\eea
which leads to a $[(\mu-\mu_{cr})\log^2(\mu-\mu_{cr})]^{-1}$-like singularity in the number susceptibility.

\noindent
A straightforward generalization of \eq{eq:nGN} to three spatial 
dimensions suggests that for $G_V=0$ the change $\Delta \nave$ in the average 
number density near the transition from the broken homogeneous to the 
inhomogeneous phase at $\mu = \mu_{cr}(T)$ should be given by 
\bea
\Delta \nave
&=&
-\frac{c\mu_{cr}^3}{\ln(\mu/\mu_{cr}-1)}\,,
\eea
with some temperature-dependent coefficient $c$.
Indeed, our numerical results are consistent with this behavior,
thus explaining the divergence of $\chi_{nn}$ at  $\mu = \mu_{cr}$.

\noindent
In contrast, for $G_V>0$ the mapping $\tilde{\mu}\to\mu$ via Eq.(\ref{eq:gapeqmueff}) leads to a qualitatively different behavior. For this case we find
\bea
\chi_{nn}
\quad=\quad
\frac{\partial n}{\partial \tilde{\mu}}/
\frac{\partial \mu}{\partial \tilde{\mu}}
\quad=\quad
\left.
\frac{
1
}{
1+2G_V\frac{\partial n}{\partial \tilde{\mu}}
}
\frac{\partial n}{\partial \tilde{\mu}}
\right\vert_{\tilde{\mu}(\mu)}
\,.
\eea
Therefore a divergence in $\partial n/\partial \mu$ at $G_V=0$ does not result in a divergent number susceptibility for $G_V>0$, but merely leads to a jump of order $1/2G_V$. This is illustrated on the right hand side of Fig.~\ref{fig:susc3d}.

\subsubsection{Finite current quark masses}

\begin{figure}
\includegraphics[width=.6\textwidth]{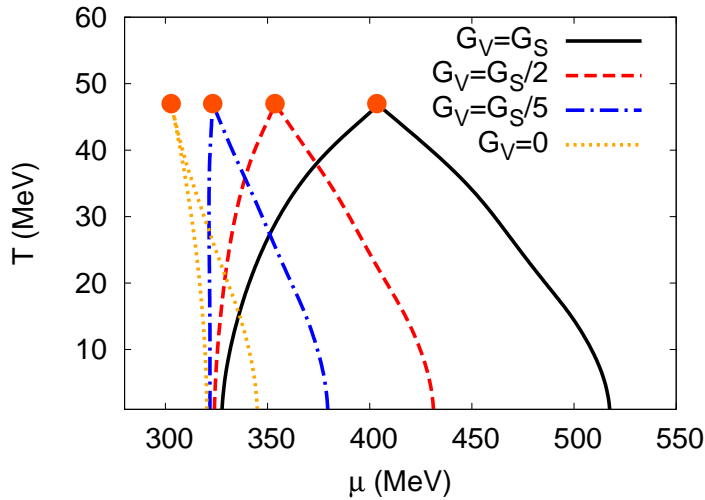}
\caption{$\mu-T$ phase diagram for different values of $G_V$ and a 
current quark mass $m = 5 \,\mathrm{MeV}$.
The dots indicate the Lifshitz points above which the second order transition 
lines join and turn into a crossover.}
\label{fig:m}
\end{figure}

\noindent
For completeness we present in Fig.~\ref{fig:m} our results for the phase
diagram at a non-vanishing current quark mass $m=5$~MeV and various values
of $G_V$ by considering \eq{eq:MzReal}. 
Since there is no exact order parameter to distinguish the homogeneous broken 
from the restored phase in this case, there are strictly speaking only
two phases -- a homogeneous and an inhomogeneous phase -- and, hence,
no Lifshitz point. Nevertheless we can easily identify the remnant of the
LP as a cusp in the phase boundary. 
For simplicity, we will call this point a Lifshitz point as well. 

\noindent
It has already been 
observed~\cite{Nickel:2009wj} that the LP shifts to smaller 
temperatures and larger chemical potentials  when increasing $m$. 
The dependence on $G_V$ however stays qualitatively the same as in the chiral 
limit: the LP is only shifted in the $\mu$-direction upon increasing $G_V$ and 
the domain of inhomogeneous phases in the $\mu-T$ diagram is stretched. 
The explanation for this behavior is identical to the one given in 
section~\ref{subsec:phasediag} for the chiral limit.

\subsection{Ginzburg-Landau expansion}
\label{sec:GL}

\noindent
In the vicinity of a second-order phase transition and in particular of a 
critical point, 
the thermodynamic potential can be studied systematically within a 
Ginzburg-Landau expansion. 
In the present case, this corresponds to an expansion of the thermodynamic
potential as an effective action in $\delta M(\x) = M(\x) - M_0$
and $\delta\tmu(\x) = \tmu(\x) - \tmu_0$ around their values 
$M(\x) = M_0$ and $\delta\tmu(\x) =\tmu_0$ in the restored phase.
For simplicity, we restrict ourselves to the chiral limit, so that 
$M_0 = 0$ and therefore $\delta M(\x) = M(\x)$.
However, unlike in the numerical studies above, we will not assume
$\tilde\mu(\x)$ to be spatially uniform. 

\noindent 
At given temperature and chemical potential the expansion then takes the
form
\bea
\Omega[M,\tilde{\mu}]
&=&
\Omega[0,\tilde{\mu}_0]
+
\frac{1}{V}\int d\x \; \Omega_{GL}(M(\x),\delta\tilde{\mu}(\x))\,,
\eea
with
\begin{alignat}{1}
&\Omega_{GL}(M(\x),\delta\tilde{\mu}(\x))
\nonumber\\
&=\;
c_{2,a}
 \vert M(\x)\vert^2
+
c_{2,b}
\delta\tilde{\mu}(\x)^2
+
c_{3,a}
\vert M(\x)\vert^2
\delta\tilde{\mu}(\x)
+
c_{3,b}
\delta\tilde{\mu}(\x)^3
\nonumber\\
&\quad+
c_{4,a}
\vert M(\x)\vert^4
+
c_{4,b}
\vert\nabla  M(\x)\vert^2
+
c_{4,c}
 \vert M(\x)\vert^2
\delta\tilde{\mu}(\x)^2
+
c_{4,d}
\delta\tilde{\mu}(\x)^4
+
c_{4,e}
(\nabla \delta\tilde{\mu}(\x))^2
+
\dots
\,,
\end{alignat}
when expanding to fourth order in $M(\x)$, $\delta\tilde{\mu}(\x)$ and 
gradients acting on these functions.
The symmetries of the theory and of the background simplify the expansion: 
linear terms vanish as we are expanding around a homogeneous solution of
the gap equations, 
odd terms in $M(\x)$ vanish by chiral symmetry and odd numbers of derivatives 
vanish due to rotational symmetry.

\noindent
Taking $M(\x)$ to be the small scale of interest, we first aim at an estimate 
for $\delta\tilde{\mu}(\x;M(\x))$ defined through the stationary constraint
\bea
\left.
\frac{
\delta \Omega
}{
\delta \delta \tmu
}
\right|_{M(\x),\delta\tilde{\mu}(\x)=\delta\tilde{\mu}(\x;M(\x))}
&=&
0
\,.
\eea
Because of the absence of a linear term in $M(\x)$
we conclude that $\delta\tmu(\x;M(\x))\sim O(|M(\x)|^2)$. 
More precisely, 
\beq
    \delta\tmu(\x;M(\x)) = -\frac{c_{3,a}}{2 c_{2,b}}|M(\x)|^2+\dots \,.
\eeq
Consequently, the expansion of 
$\Omega_{GL}(M(\x))\equiv \Omega_{GL}(M(\x),\delta\tmu(\x;M(\x)))$ to fourth 
order is given by
\bea
\Omega_{GL}(M(\x))
&=&
\Omega_{GL}[0,\tmu_0]
+
c_{2,a}\vert M(\x)\vert^2
+
\left(
c_{4,a}
-
\frac{c_{3,a}^2}{4c_{2,b}}
\right)
\vert M(\x)\vert^4
+
c_{4,b}
\vert\nabla  M(\x)\vert^2
+
\dots
\,,
\eea
which allows us to determine the Lifshitz and the critical points from the 
GL coefficients.
The latter is characterized by vanishing quadratic and quartic mass terms, 
while at the former the quadratic term and the gradient term are zero:  
\bea
\label{eq:CPLP}
&&\text{LP: }\quad 0\,=\,c_{2,a}\,=\,c_{4,b}
\,,
\nonumber\\
&&\text{CP: }\quad 0\,=\,c_{2,a}\,=\,c_{4,a}
-
\frac{c_{3,a}^2}{4c_{2,b}}
\,.
\eea
As outlined in Ref.~\cite{Nickel:2009ke} for the NJL model without vector
interactions, it is a straightforward exercise to work out the explicit form 
of the GL coefficients.
As obvious from \eqs{eq:Omega1} and (\ref{eq:hamiltonian0}),
the kinetic part of the thermodynamic potential, $\Omega_{\rm kinetic}$,
and, hence, its contributions to the GL coefficients
depend on $G_V$ only indirectly through $\tilde{\mu}_0$ and $T$.
The only explicit dependence on $G_V$ therefore originates from 
$\Omega_{\rm cond}$ and only affects $c_{2,b}$.
For this reason the Lifshitz point as a function of $\tilde{\mu}_0$ and $T$ is 
independent of $G_V$, in agreement with our findings in 
section~\ref{subsec:numGv}. 

\noindent
In contrast there is an explicit dependence of the location of the critical 
point on $G_V$ through $c_{2,b}$. One finds
\beq
    c_{2,b} = 
    -\frac{1}{4G_V} 
    - N_c\left(\frac{\tmu_0^2}{\pi^2} + \frac{T^2}{3}\right)\,, 
\eeq 
where the first term on the right hand side is due to $\Omega_{\rm cond}$,
while the second term is due to $\Omega_{\rm kinetic}$. 
Since we are expanding around a homogeneous restored solution, the latter
is just given by the corresponding term in an ideal Fermi gas at 
temperature $T$ and chemical potential $\tmu_0$. Here we have 
neglected the contributions from the regulator terms,
which would arise in our specific model. However, these terms are small in 
the region of interest and therefore do not lead to qualitative changes.
Hence, $1/c_{2,b}$ vanishes for $G_V=0$ and decreases monotonously with
increasing positive values of $G_V$. 
Furthermore $c_{4,a}$ typically decreases when increasing $\tilde{\mu}_0$ or 
decreasing $T$ in the vicinity of the critical point.
Put together, we conclude from Eq.~(\ref{eq:CPLP}) that the CP 
moves to smaller temperatures upon increasing $G_V$.
Moreover,
discarding possible issues related to the regularization of the UV-divergent 
vacuum contribution to the thermodynamic potential\footnote{For a 
renormalizable theory divergent GL coefficients are subject to 
renormalization; the present discussion is consistent for a regularization 
scheme acting on the energy spectrum as applied in this work.},
we find $c_{4,a}= c_{4,b}$, as in the case without vector 
interaction~\cite{Nickel:2009ke}. 
For $G_V=0$ we therefore recover the result of Ref.~\cite{Nickel:2009ke}
that the LP and the CP coincide.

\noindent
Since $\delta\tmu(\x)\sim O(M(\x)^2)$ we can also conclude that the 
thermodynamic potential expanded around $\tilde{\mu}_0$ to order $O(|M(\x)|^2)$ 
and arbitrary gradients coincides with that of the model for $G_V=0$
upon replacing $\mu\rightarrow\tilde{\mu}_0$. As a result, the second-order 
phase transition from any inhomogeneous to the chirally restored phase, being 
triggered by these contributions, is only depending on $G_V$ through 
$\tilde{\mu}_0$, as it was already obtained in section~\ref{subsec:numGv} 
by applying further truncations.

\noindent
On the other hand, the present analysis is not applicable to the transition 
from the homogeneous broken to inhomogeneous phase, where the mass function
is not related to a small parameter, when we go away from the LP. 
As we have argued earlier, even the order of the phase transition may
change in this regime when vector interactions are included.

\noindent
Qualitatively the same properties as discussed here for the NJL model with a vector 
interaction also show up in the Gross-Neveu model with a Thirring interaction 
(GNT model) at leading order in the large $N$-expansion. Since the model is 
renormalizable and therefore much cleaner and easier to handle, we include a 
discussion of its phase diagram in Appendix~\ref{app:GNT}.

\section{Effects of Polyakov loop dynamics}
\label{sec:Polyakov}

\subsection{Inclusion of the Polyakov loop}
\label{subsec:Polyakov}

\noindent
In order to mimic features of confinement, in particular to suppress the 
contribution of free constituent quarks in the confined phase,
and in order to include gluonic contributions to the pressure, 
the NJL model can be coupled to an effective description of the Polyakov 
loop~\cite{Ratti:2005jh,Fukushima:2003fw}. The resulting model is known
as the Polyakov-loop extended Nambu--Jona-lasinio (PNJL) model.

\noindent
The Polyakov loop is defined by
\beq
    L(\x) = {\cal P} \exp\left[i\int_0^{1/T} d\tau\, A_4(\tau,\x) \right]\,,
\label{eq:Poly}
\eeq
where $A_4(\tau,\x) = iA_0(t=-i\tau,\x)$ is the temporal part of a gauge 
field $A_\mu = g A_\mu^a \frac{\lambda^a}{2}$ at imaginary time.
In pure Yang-Mills theory, the traced expectation values of $L$ 
and its hermitean conjugate, 
\beq
     \ell=\frac{1}{N_c}\langle{\rm Tr}_c L\rangle 
\,,
     \qquad
     \bar\ell=\frac{1}{N_c}\langle{\rm Tr}_c L^\dagger\rangle
\,,
\label{eq:llbar}
\eeq
can be related to the free energies of a static 
quark or antiquark,
$\ell \sim e^{-F_q/T}$, 
$\bar\ell \sim e^{-F_{\bar q}/T}$~\cite{McLerran:1981pb,Karsch:1985cb},
and are therefore order parameters for the confinement-deconfinement 
transition.

\noindent
To include the Polyakov-loop dynamics in the NJL model,
the quarks are minimally coupled to a background gauge field, 
$\partial_\mu \rightarrow D_\mu = \partial_\mu + iA_0 \delta_{\mu 0}$.
Furthermore, a local potential $\mathcal{U}(\ell,\bar{\ell})$ is added to 
the thermodynamic potential,
which is essentially constructed to reproduce ab-initio results of pure 
Yang-Mills theory at finite temperature~\cite{Ratti:2005jh,Roessner:2006xn}.
In the most simple approach, the gauge field is taken to be a 
space-time independent mean field $A_4$, 
and the effect of the covariant derivative in the kinetic part
amounts to the replacement~\cite{Fukushima:2003fw} 
\bea
\label{eq:fPNJL}
f_{\rm thermal}(E)
&\rightarrow&
f_{\rm thermal, PNJL}(E)
=
T\ln\left(
1 + \mathrm{e}^{-3(E-\mu)/T}
  +3\,\ell\,\mathrm{e}^{-(E-\mu)/T}
   +3\,\lb\,\mathrm{e}^{-2(E-\mu)/T}\right)
   \nonumber\\
&&
\phantom{f_{\rm thermal, PNJL}(E)}
+T\ln\left(
1 + \mathrm{e}^{-3(E+\mu)/T} 
 +3\,\lb\,\mathrm{e}^{-(E+\mu)/T}
   +3\,\ell\,\mathrm{e}^{-2(E+\mu)/T}\right) 
\,.
\eea
However, as can be seen from \eqs{eq:Poly} and (\ref{eq:llbar}),
a constant mean field $A_4$ would always result in 
complex conjugate values for $\ell$ and $\bar\ell$.
This should be considered as an artifact, since their interpretation 
as being related to the free energies of quarks and antiquarks
means that $\ell$ and $\bar\ell$ should be real and, 
at finite chemical potential, different from each other. 
In order to by-pass this problem, we therefore follow the viewpoint of
Ref.~\cite{Fukushima:2008} that, after the replacement (\ref{eq:fPNJL}), 
$\ell$ and $\bar\ell$ should be treated as independent real mean fields,
rather than the components of $A_4$.  
This is also consistent with the fact that the potential 
$\mathcal{U}$ which is added to the thermodynamic potential
is given in terms of $\ell$ and $\bar\ell$ as well.
For simplicity we take Fukushima's parametrization~\cite{Fukushima:2008},
\bea
\label{eq:Ullbar}
\mathcal{U}(\ell,\bar{\ell})
&=&
-b\,T \left(
54\,
\mathrm{e}^{-a/T}\,\ell\,\lb \, +\log\bigl[1-6\,\ell\,\lb -3(\,\ell\,\lb\,)^2+4(\,\ell^3+\lb\,^3)\bigr] \right)
\eea
with two parameters $a$ and $b$. Other prescriptions, like the
polynomial~\cite{Ratti:2005jh} or the logarithmic~\cite{Roessner:2006xn} 
potential, would lead to very similar results.

\noindent
When dealing with inhomogeneous phases, $\ell$ and $\bar{\ell}$ are
naturally expected to be spatially dependent, presumably following the
density profile in some way. 
Nevertheless, similar to the treatment of $\tmu$ in the previous section,
we will assume spatially independent values of $\ell$ and $\bar{\ell}$,
even in the inhomogeneous phase.
This is not only to keep the technical side of the calculation trackable,
but also the assumptions made in order to derive (\ref{eq:fPNJL}) and the 
unknown kinetic contributions to \eq{eq:Ullbar} suggest such a conservative 
approach as a first step.

\noindent
To summarize, we obtain a thermodynamic potential
\bea
\Omega_{\rm PNJL}
&=&
\left.\Omega_{\rm kinetic}\right\vert_{f_{\rm thermal}\rightarrow f_{\rm thermal,PNJL}}
+
\Omega_{\rm cond}
+
\mathcal{U}(\ell,\bar{\ell})
\,,
\eea
with $\Omega_{\rm kinetic}$ and $\Omega_{\rm cond}$ given in 
\eqs{eq:Omega1}-(\ref{eq:Omegaf2}), that additionally needs to be extremized 
in $\ell$ and $\bar\ell$.

\subsection{Numerical results}
\label{subsec:Polyakov_numerics}

\noindent
Since we consider the NJL model with $G_V=m=0$ as our starting point, 
we shall limit ourselves to this case when studying the role of the Polyakov 
loop.
For the Polyakov-loop potential, we adopt the parameters of 
Ref.~\cite{Fukushima:2008},
$a=664\,\mathrm{MeV}$ and $b=7.55 \cdot 10^6\, \mathrm{MeV}^3$.
The parameter $a$ was fixed by the condition that for pure gluo-dynamics 
the phase transition takes place at $T = 270$~MeV, while $b$ was chosen to
have a crossover around $T = 200$~MeV at $\mu = 0$ when quarks are included.
Since we are mainly interested in the qualitative effect of the Polyakov
loop, we did not perform a refit of $b$ within our regularization scheme.
We checked, however, that this parameter choice gives reasonable results
for the behavior of the order parameters at $\mu = 0$. 

\begin{figure}
\begin{center}
\includegraphics[width=.6\textwidth]{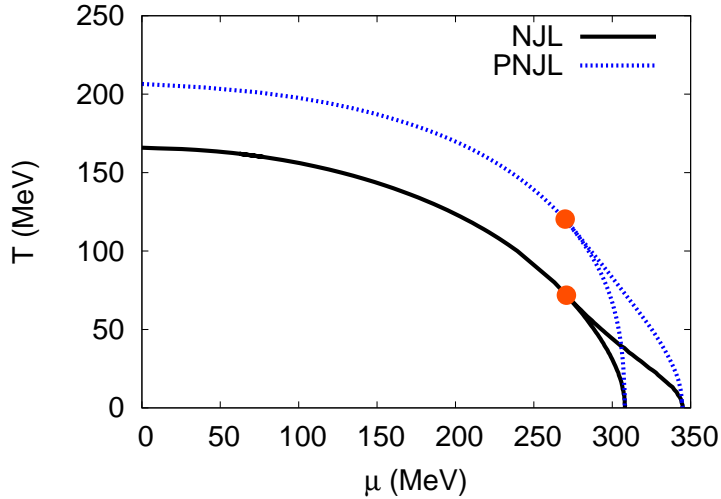}
\caption{Phase diagram of the NJL (solid line) and PNJL (dashed line) model allowing for one-dimensional spatial modulations of the order parameter.}
\label{fig:pdpnjl}
\end{center}
\end{figure}

\noindent
In Fig.~\ref{fig:pdpnjl} we compare the phase diagrams for the NJL model with
that of the PNJL model, allowing for phases with a one-dimensional solitonic 
modulation in both cases.
Aside from a general stretching in the $T$-direction, which is well known 
from studies of homogeneous phases and easily explained by the replacement 
(\ref{eq:fPNJL}), the Polyakov loop has no effect on the qualitative structure 
of the phase diagram. In particular, the critical point at $G_V=0$ still 
coincides with the Lifshitz point.  

\begin{figure}
\begin{center}
\includegraphics[width=.6\textwidth]{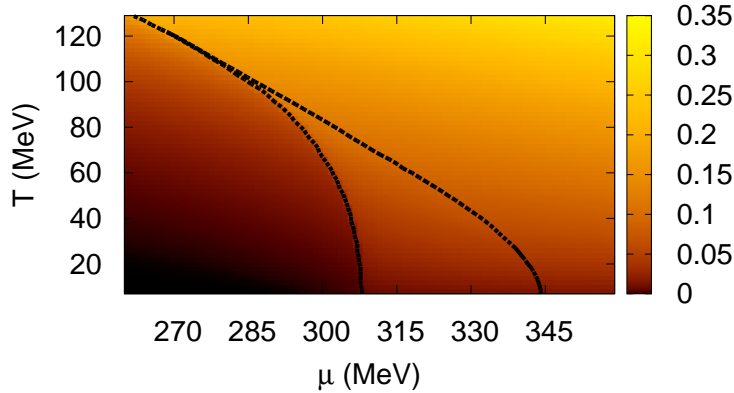}
\caption{Polyakov loop expectation value $\ell$ in the $\mu$-$T$ plane 
in the vicinity of the inhomogeneous phase.}
\label{fig:pdpnjl2}
\end{center}
\end{figure}

\noindent
In Fig.~\ref{fig:pdpnjl2} the value of $\ell$ is presented via color
coding in the region of the phase diagram where the inhomogeneous 
phase is favored. We find that $\ell$ and $\bar\ell$ are rather small in 
the entire inhomogeneous phase, reaching their maximum values
$\ell \approx 0.15$ and $\bar\ell \approx 0.2$ near the LP.
In this context we should recall that at vanishing temperature the Polyakov-loop 
dynamics decouples completely from the quark sector due to the way the PNJL model 
is constructed. 
As a consequence, $\ell=\bar\ell=0$ at $T=0$, independent of the density. 
While it is unclear whether this feature of the model is realistic,
it means that our assumption of space-independent Polyakov-loop expectation 
values cannot have a large effect. 
Even if $\ell$ and $\bar\ell$ followed the density profile, 
the results would not be very different, because at low temperatures
their values are very small anyway, whereas at higher temperatures the 
density differences get washed out.

\section{Discussion}
\label{sec:discussion}

\noindent
In this work we have analyzed the role of the isoscalar-vector interaction and the dynamics of the Polyakov loop on recently discussed inhomogeneous ground states in the phase diagram of the NJL model.
Mainly for technical reasons we thereby limited ourselves to inhomogeneous phases with a one-dimensional modulation, explicitly to domain-wall soliton lattices and for comparison to chiral spirals. This allowed us to exploit the knowledge obtained for lower dimensional models in our study.

\noindent
Our previous study in absence of vector interactions has led us to the finding that the critical point in the mean-field phase diagram of the NJL model actually coincides with a Lifshitz point when allowing for the possibility of inhomogeneous phases. The first order phase transition is then absent in the phase diagram, since an inhomogeneous phase is energetically preferred in its domain.
When extending our model, we find that a repulsive vector interaction leads to significant qualitative effects:
In contrast to the critical point when limiting to homogeneous phases, the Lifshitz point is not shifted towards smaller temperatures when increasing the strength of the vector interaction, but remains at the same temperature and density. Moreover, the domain of inhomogeneous ground states in the $\mu-T$ phase diagram increases.
Since the Lifshitz point therefore no longer coincides with the critical point, the critical behavior in its vicinity and also near the phase transition lines to the inhomogeneous phase changes. This is underlined by the fact that the number susceptibilities remain finite, i.e., there are no singularities.

\noindent
Our investigation of this part is complemented by an extensive numerical study, including the determination of density profiles in the inhomogeneous phase, the phase diagram when only allowing for chiral spirals, the evaluation of number susceptibilities and the incorporation of finite current quark masses. Furthermore, we performed a generalized Ginzburg-Landau expansion for elaborating the dependence of Lifshitz and critical point on the vector interaction and we mapped out the phase diagram of the $1+1$-dimensional GN model with a Thirring interaction.
Part of this comprehensive study is also motivated in order to back up an approximation employed within our mean-field calculation, namely to use the spatially averaged number density when evaluating the thermodynamic potential.

\noindent
Probably less spectacular but nevertheless worth checking is the behavior of our model when the quarks are coupled to the Polyakov loop in order to suppress their contribution to the thermodynamic potential in chirally broken phases, thus mimicking confinement.
In the absence of a vector interaction this coupling, at least in the employed approximation, does not lead to a separation of Lifshitz and critical point. Consequently the $\mu-T$ phase diagram is not modified qualitatively and, similar as in the case of homogeneous phases, is only stretched in the temperature direction.

\noindent
Our analysis has shown that the region where solitonic
modulations of the chiral order parameter are favored over homogeneous
phases increases when considering natural extensions of the two-flavor
NJL model. Although being much more tractable on the technical side, it
is worth noting that strictly speaking one-dimensional modulations of
the order parameter are washed out by thermal fluctuations at finite
temperature \cite{Landau:1969, Baym:1982}. This statement refers
to the fact that the modulation is not rigid, but fluctuates locally.
The system's spatially modulated nature is however still encoded in the
behavior of long range correlations.
To get past this kind of issue, it would anyway be of great interest to
extend the present analysis to inhomogeneous phases with higher
dimensional modulations. This should be possible in a more numerical
approach as outlined in Ref.~\cite{Nickel:2008ng} in the context of inhomogeneous
color-superconducting phases. The most interesting question here is
whether the transition from the inhomogeneous to the restored phase can
turn first order for more complex modulations at low enough
temperatures. E.g. for color-superconducting phases, this has been
suggested in Ref.~\cite{Bowers:2002xr}, although the applied expansion did not allow for
a conclusive answer.

\section*{Acknowledgments}
\noindent
The authors would like to thank B.~Friman, K.~Fukushima, 
C.~Tsallis, and J.~Wambach for helpful comments and discussions. 
This work was partially supported by the Department of Energy (DOE) under 
grant numbers DE-FG02-00ER41132 and DE-FG0205ER41360, 
by the German Research Foundation (DFG) under grant number Ni 1191/1-1, 
by the Helmholtz research school for Quark Matter studies (H-QM),
and by the Helmholtz Alliance EMMI.

\appendix

\section{The Gross-Neveu model with a Thirring interaction}
\label{app:GNT}
\noindent
To back up our results related to the vector interaction we briefly want to 
discuss its effect in the Gross-Neveu model with a Thirring interaction 
(GNT model).
Unlike the NJL model, the GNT model is renormalizable and therefore does not
suffer from regularization artifacts. 
The relevant Lagrangian is given by
\bea
\mathcal{L}
&=&
 \bar{\psi}\left(i\gamma^\mu \partial_\mu -m\right)\psi +
\frac{G_S}{2N}\left(\bar{\psi}\psi\right)^2
-
\frac{G_V}{2N}  \left(\bar{\psi}\gamma^\mu\psi\right)^2
\,,
\eea
where  $\psi$ is a $2N$-dimensional spinor for $N$ species in $1+1$ dimensions 
and $\gamma^\mu$ can be chosen as $\gamma^0=\sigma^1$, $\gamma^1=-i\sigma^2$, 
$\sigma^i$ being the usual Pauli matrices.

\noindent
In the large $N$-limit the mean-field approximation becomes 
exact.\footnote{Note that this limit does not commute with the thermodynamic 
limit.} Furthermore, 
since the model is renormalizable, all divergencies can be absorbed into a 
finite number of contact terms.
The thermodynamic potential is given by
\bea
\label{eq:OGNT}
\Omega_{\rm GNT}/N
&=&
\Omega_{\rm kinetic}/N + \Omega_{\rm cond}/N
\,,
\nonumber\\
\Omega_{\rm kinetic}/N
&=&
-\frac{T}{L} \sum_n {\rm Tr}_{D,L} {\rm Log} \left(\frac{1}{T}(i\omega_n+F)\right)
\,,
\nonumber\\
\Omega_{\rm cond}/N
&=&
\frac{1}{L}\int dz \left(\frac{(M(z)-m)^2}{2G_S}-\frac{(\tilde{\mu}(z)-\mu)^2}{2G_V}\right)
\,,
\eea
where we introduced a similar notation as in the case of the NJL model: 
$L$ is the periodicity, the spatial coordinate is labelled $z$, 
$M(z)=m-G_S\langle\bar{\psi}\psi\rangle/N$, 
$\tilde{\mu}(z)=\mu- G_V\langle\bar{\psi}\gamma^0\psi\rangle/N$ and
\bea
F
&=&
-i\gamma^0\gamma^1\partial_z + \gamma^0 M(z) -\tilde{\mu}(z)
\,.
\eea
For this setup we first investigate the phase diagram of homogeneous phases, 
in particular the role of $G_V$, and subsequently also inhomogeneous phases 
by means of a Ginzburg-Landau expansion.

\noindent
For homogeneous phases the eigenvalue spectrum of $F$ can be labelled by the momenta and it is straightforward to obtain
\bea
\Omega_{\rm GNT}/N
&=&
\underbrace{
-2\int_0^\Lambda \!\frac{dp}{2\pi} 
E_p
+
\frac{(M-m)^2}{2G_S}-\frac{(\tilde{\mu}-\mu)^2}{2G_V}
}_{\equiv \Omega_{\rm vacuum}}
\underbrace{
-
2T\int_0^\infty \!\frac{dp}{2\pi} 
\sum_{s=\pm}
\ln
\left(1+e^{-\frac{E_p+s \tilde{\mu}}{T}}\right)
}_{\equiv \Omega_{\rm thermal}}
\,,
\eea
where $E_p=\sqrt{p^2+M^2}$.
For $\Omega_{\rm vacuum}$ we introduced a sharp momentum cutoff $\Lambda$ as 
the integral is divergent. Note, however, that for a renormalizable theory
- in contrast to the NJL model - the regularization scheme is only introduced 
intermediately and does not affect the final results. We also note that the 
argument of the vacuum contribution should in principle contain the 
eigenvalues $\pm E_p-\tilde{\mu}+\mu$,\footnote{This is related to the fact 
that, like the constituent mass $M$, 
the density  $\langle\bar{\psi}\gamma^0\psi\rangle \propto \mu - \tmu$ is 
undetermined before the gap equations are solved.}
but for a sharp momentum cutoff this 
does not affect the renormalization procedure at large momenta.
Concerning the ultraviolet behavior, we observe that the quadratic divergency 
can be absorbed into an order parameter independent constant, a linear and a 
quadratic term in $M$, corresponding to a renormalization of the overall 
pressure at $M=0$, the fermion mass $m$ and the coupling $G_S$.
Requiring the thermodynamic potential to vanish at 
$M= \langle\bar{\psi}\gamma^0\psi\rangle=0$ 
and to have a minimum at $M_0$, we then get
\bea
\Omega_{\rm vacuum}/N
&=&
\frac{1}{4\pi} M^2 \left(-1+2 \ln\frac{M}{M_0}\right)
+
\frac{c M \left(M-2M_0\right)}{2 M_0}
-
\frac{(\tilde{\mu}-\mu)^2}{2G_V}
\,,
\eea
where $c$ is proportional to $m$ at finite $\Lambda$ and not relevant here as we shall 
consider the chiral limit $c=0$ in the following.
$G_V$ remains finite and can be assumed to have its renormalized value. 
In order to discuss the phase diagram all that is left is to extremize the potential in $M$ and $\tilde{\mu}$.
For simplicity we will focus the discussion on phase transitions and the 
location of the critical point.

\begin{figure}
\begin{center}
\includegraphics[width=.6\textwidth]{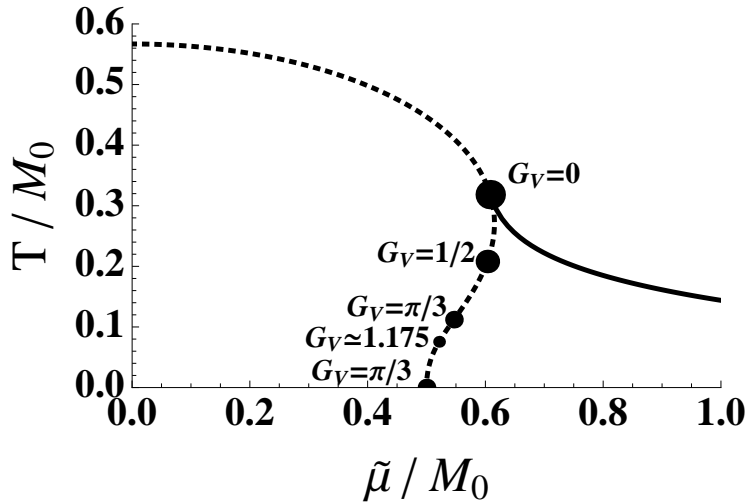}
\caption{
Phase diagram of the massless GNT model: The dotted line corresponds to 
$\alpha_{2,0}=0$, the dots indicate the critical points when limiting to 
homogeneous phases, i.e., in addition $\beta_4=0$. 
At $G_V=\pi/3$ an additional CP emerges on the $\tilde{\mu}$-axis, 
which merges with the other CP at $G_{V,{\rm max}}\simeq1.175$. 
The solid line shows the transition from the chiral crystalline to the 
restored phase, ending at the Lifshitz point. 
The latter agrees with the CP at $G_V=0$.
}
\label{fig:plGNT}
\end{center}
\end{figure}

\noindent
Addressing phase transitions between homogeneous phases, 
we first expand the thermodynamic potential in the constituent mass
around $M=0$. Similar to the GN model~\cite{Boehmer:2007ea} we find
\bea
\Omega_{\rm GNT}/N
&=&
\sum_{n\geq0} \alpha_{2n}(\tilde{\mu})M^{2n}
\eea
with
\bea
\alpha_0(\tmu)
&=&
-\frac{\pi T^2}{6}
-\frac{ \tmu^2}{2\pi}
-\frac{(\tilde{\mu}-\mu)^2}{2G_V}
\,,
\nonumber\\
\alpha_{2}(\tmu)
&=&
\frac{1}{2\pi}\left(\ln\left(\frac{4\pi T}{M_0}\right)+ {\rm Re}\,\psi(z) \right)
\,,
\nonumber\\
\alpha_{2n}(\tmu)
&=&
-\frac{(-1)^n}{2^{2n-1}(n-1)(2n-4)!!(2n)!!\pi^{2n-1}T^{2n-2}}{\rm Re}\,\psi(2n-2,z)
\,,\quad n>1\,,
\eea
where we used the di- and polygamma functions and 
$z=\frac{1}{2}+\frac{i\tilde{\mu}}{2\pi T}$.
Note that only $\alpha_{2}$ is affected by the renormalization procedure.
From the additional constraint 
${\partial \Omega_{\rm GNT}}/{\partial \tilde{\mu}}=0$ we can then infer the 
value $\tilde{\mu}_0=\frac{\pi\mu}{\pi+G_V}$ of the renormalized chemical 
potential at $M=0$ and furthermore expand in 
$\delta\tilde{\mu} = \tilde{\mu}-\tilde{\mu}_0$.
For the purpose of our discussion, we can then limit ourselves to
\bea
\Omega_{\rm GNT}/N
&=&
\Omega_{\rm unbroken}
+
\alpha_{2,0}M^2
+
\alpha_{4,0}M^4
+
\alpha_{2,1}M^2\delta\tilde{\mu}
+
\alpha_{0,2}\delta\tilde{\mu}^2
+
O(M^6,\delta\tilde{\mu}^3,M^2\delta\tilde{\mu}^2,M^4\delta\tilde{\mu})
\,,
\nonumber\\
\eea
where $\alpha_{n,0} = \alpha_n|_{\tmu=\tmu_0}$
and
\bea
\Omega_{\rm unbroken}
\equiv 
\alpha_{0,0}
&=&
-\frac{\pi T^2}{6}
-\frac{ \mu^2}{2(\pi+G_V)}
\,,
\nonumber\\
\alpha_{2,1}
&=&
-
\frac{1}{4\pi^2 T} {\rm Im}\,\psi(1,z)|_{\tmu=\tmu_0} 
\,,
\nonumber\\
\alpha_{0,2}
&=&
-\frac{\pi+G_V}{4\pi G_V}
\,.
\eea
We conclude that $\delta\tilde{\mu}=-\frac{\alpha_{2,1}}{2\alpha_{0,2}}M^2+O(M^3)$ and arrive at an expansion of the thermodynamic potential only depending on $M$,
\bea
\Omega_{\rm GNT}/N
&=&
\Omega_{\rm unbroken}
+
\alpha_{2,0}M^2
+
\beta_{4}M^4
+
O(M^6)
\,,
\eea
where
\bea
\beta_4
&=&
\alpha_{4,0}-\frac{\alpha_{2,1}^2}{4\alpha_{0,2}}
\,.
\eea
The critical point is then given by $\alpha_{2,0}=\beta_4=0$. 
As a simple exercise we can evaluate this condition at zero temperature,
for which one finds $\ln(2\tilde{\mu}_0/M_0)=-1+\frac{3}{\pi}G_V=0$. 
At $G_V=\pi/3$ we have therefore a critical point at $T=0$ and $\tmu=2M_0/3$. 
However, as illustrated in Fig.(\ref{fig:plGNT}), it turns out that
this is a ``new'' CP, which moves upwards in temperature upon increasing 
$G_V$. 
On the other hand, the ``old'' CP, existing already at $G_V=0$, 
moves downwards, similar to the NJL model.
Eventually, at $G_{V}\simeq1.175$, both points merge and 
no first-order phase transition is left at higher values of $G_V$.

\noindent
Turning to inhomogeneous phases, we can perform a generalized Ginzburg-Landau 
expansion. As in the NJL-model case, discussed in Sec.~\ref{sec:GL},
this corresponds to an expansion of (\ref{eq:OGNT}) in $M(z)$ and 
$\delta\tilde{\mu}(z)=\tilde{\mu}(z)-\tilde{\mu}_0$, combined with a 
derivative expansion in order to obtain a local functional.
The technology has been worked out in great detail for the GN 
model~\cite{Thies:2006ti}
and equally works for the GNT model.
Since a $M(z)\delta\tilde{\mu}(z)$-term is forbidden by $\mathbb{Z}_2$ symmetry, we have in general
$\delta\tilde{\mu}(z)\sim M(z)^2$.
Furthermore, by treating derivatives to be of order $O(M(z))$ we get to fourth order
\bea
\Omega_{\rm GNT}/N
&=&
\Omega_{\rm unbroken}
+
\frac{1}{L}\int\!dz\Big(
\alpha_{2,0}M(z)^2
+
\alpha_{4,0}(M(z)^4+M'(z)^2)
+
\alpha_{2,1}M^2\delta\tilde{\mu}(z)
\nonumber\\&&\hspace{4cm}
+
\alpha_{0,2}\delta\tilde{\mu}(z)^2
\Big)
+
\dots
\eea
and observe that the prefactors of the $M(z)^2$ and $M'(z)^2$ terms are those 
obtained for the GN model upon replacing $\mu\rightarrow\tilde{\mu}_0$. From 
this we conclude that the location of the Lifshitz point in the 
$\tilde{\mu}-T$ diagram is the same as in the GN model. Consequently, 
since the density there is directly given through $\tilde{\mu}_0$, 
$G_V$ does not affect the Lifshitz point in the $\nave-T$ diagram, 
while in the $\mu-T$ diagram the LP is shifted 
by $G_V\langle\bar{\psi}\gamma^0\psi\rangle/N$ in the $\mu$-direction.

\noindent
Another property that is inherited from the GN model is the second-order transition line from the chiral crystalline to the restored phase.
Since in this case the magnitude of the order parameter vanishes continuously while the modulation stays finite, we have to consider the GL expansion to order $O(M(z)^2)$ and in principle to all orders of gradients.
However, since $\delta\tilde{\mu}(z)\sim M(z)^2$, $\delta\tilde{\mu}$ does not enter in this analysis and we have to have the same result as in the GN model. For the latter this transition line has been determined in Refs.~\cite{Schnetz:2004vr,Thies:2006ti} and is also depicted in Fig.(\ref{fig:plGNT}).


\end{document}